*Optimisation of Activator Solutions for Geopolymer Synthesis: Thermochemical Stability, Sequencing, and Standardisation*

Ramon Skane[a,b,c], Franca Jones [d], Arie van Riessen[b,c], Evan Jamieson[b,c], Xiao Sun[b], William D.A. Rickard[b]

[a] Corresponding Author, ramonskane@hotmail.com

[b] John de Laeter Centre, Curtin University, PO Box U 1987, Perth, WA 6102, Australia

[c] Future Battery Industries Cooperative Research Centre, The Hub Technology Park, Perth, WA 6102, Australia

[d] School of Molecular and Life Science, Curtin University, PO Box U 1987, Perth, WA 6102, Australia

## Highlights

1. Experimentally validated model quantifies thermochemical stability in activator solutions
2. Solubility model predicts time-temperature windows for optimised solution stability
3. Higher temperatures enhance thermochemical stability, cooling increases precipitation risk
4. Feedstock sequencing affects stability: $H_2O \rightarrow NaOH \rightarrow$ silicate is preferred
5. Model enables reproducible synthesis and supports SOPs for geopolymer mixing

## Keywords

abstract
**Abstract**

Geopolymers present a sustainable alternative to conventional binders: however, their commercial viability is hindered by a lack of standardised methods for preparing stabile activator solutions – alkaline feedstocks critical to geopolymer synthesis. This study presents a combined experimental and modelling approach to evaluate the thermochemical stability, solubility constraints, and silica speciation behaviour of sodium silicate-based activators. Using quantitative $^{29}$Si NMR analysis, thermodynamic stability and three-dimensional solubility modelling, this research identifies optimal preparation conditions that minimise irreversible precipitation risks and optimises mixing periods. Key findings indicate that higher solution temperatures associated with optimised activator solution preparation were found to enhance thermochemical stability and reactivity, while cooling increased viscosity and the likelihood of unstable solution behaviour, which may necessitate discarding. The order in which feedstocks are combined directly affects whether the solution becomes unstable, with an optimal sequence of water → alkali-hydroxide → soluble silicate found to ensure greater process reliability. A predictive model and accompanying visual tools enable practitioners to assess solution viability and define stability windows by quantifying initial and final/unstable periods and temperatures based on feedstock composition and solution temperature. These results contribute to improved reproducibility and quality control in geopolymer research and represent a step toward developing standard operating procedures for activator solution synthesis.

# 1 Introduction

Geopolymers are amorphous polymeric materials formed through the synthesis of reactive aluminosilicate precursors with an 'activator' solution. This reaction initiates a dissolution–gelation–polycondensation process that results in the formation of cross-linked inorganic 'geo'-polymer networks [1, 2]. Geopolymers are increasingly studied as alternatives to conventional Ordinary Portland Cement (OPC), primarily due to their potential for low embodied carbon production, by-product utilisation, and high-performance properties. Numerous case studies using geopolymer binders have shown high performance in both research and industrial contexts, achieving compressive strengths exceeding 50 MPa [3, 4], superior thermal and fire resistance [5], and up to 95% lower embodied greenhouse gas emissions (GHGs) than OPC, depending on the system used [6, 7, 8]. They also enable industrial symbiosis through the utilisation of aluminosilicate-rich by-products such as fly ash, slag and kaolin derivatives [9, 10]. OPC production and its unsustainable use of virgin materials is responsible for an estimated 7% of global greenhouse gas (GHG) emissions, positioning geopolymers as a promising pathway for decarbonising the construction sector [11, 12].

Despite significant research growth [13], the field continues to face challenges regarding the standardisation of material inputs, processing conditions, and synthesis protocols [14]. Studies often differ widely in terms of precursor characterisation, activator formulation, overall mix designing and mixing methods, leading to inconsistencies in reported properties even when similar raw materials are used [15, 16]. Further complications arise from unaccounted thermal and kinetic effects during mixing and curing, which directly influence the reactivity of activator solutions and the structure of the final geopolymer product [14, 17].

In this work we address these challenges by focusing on the chemical optimisation and process control of geopolymer activator solutions. Specifically, we examine the influence of synthesis sequencing, thermochemical stability, and mix optimisation on activator performance, with the goal of progressing toward standardised, reproducible, and optimised geopolymer systems.

## 1.1 Soluble Na-Silicate Chemistry

The complexity of geopolymer activator solutions stems largely from the intricate physiochemistry of soluble silica [18]. Despite decades of study and widespread commercial use in the form of alkaline sodium and potassium silicates across industries (e.g. construction, water treatment, agriculture), silica chemistry remains "not fully understood" [19, 20]. Proficiency in this subfield supports more informed feedstock selection, formulation, and evaluation of geopolymer products. An illustration showcasing the complex chemistry of sodium silicate is shown in Figure 1, whereby numerous silicate species of differing geometries, sodium ions and colloidal silica particles exist in a dynamically heterogeneous and complex equilibrium with each other.

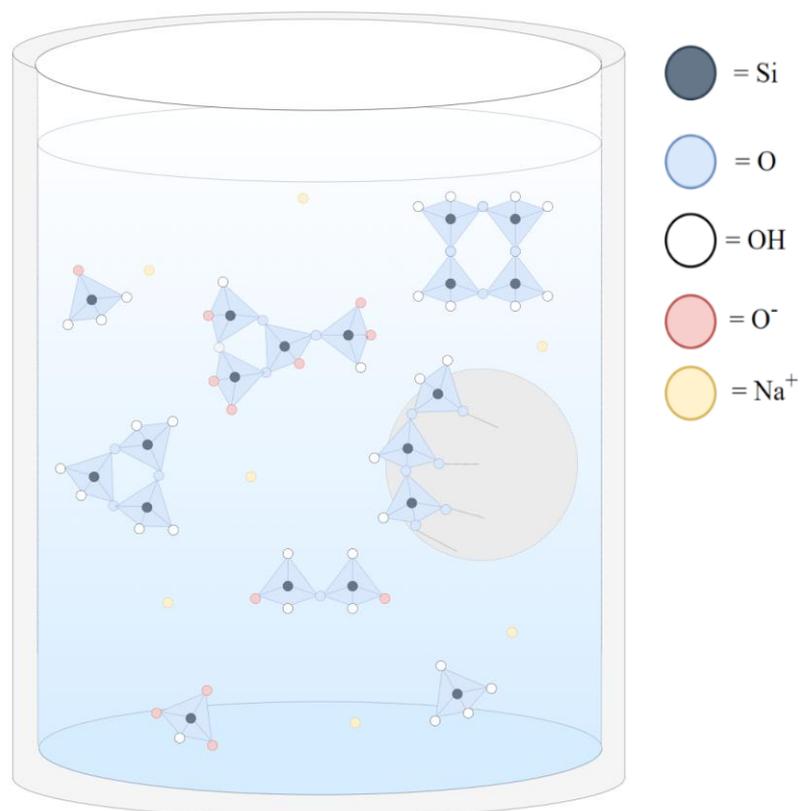

**Figure 1:** An illustration of the complex composition and geometries of chemical analytes in sodium silicate solutions, comprising various silicate species, sodium ions and a colloidal silica particle (circled in grey).

The most widely acceptable analytical method in literature for quantifying soluble silicates at the molecular level is by measuring the silica speciation within (usually alkaline) solutions. Quantification of silica speciation is used to characterise the different structural and chemical forms of soluble silicate to predict reactivity and surface chemistry, a pivotal application involving the use of siliceous precursors in producing composite materials [21, 18, 17, 22]. In describing silica speciation chemistry, most literature sources apply reference to the $Q^n$ site notation popularised by Engelhardt et al. [23] whereby a centralised tetrahedrally connected silica atom is connected to other silica tetrahedra by *n* bridging and (*4-n*) non-bridging oxygen atoms. In this notation, *n* can range between 0-4 where $Q^0$ and $Q^3$ represent monomeric and tetrameric silicates, respectively, each with different reactivities, dynamics and varying structural geometries [21, 24].

As the characteristic degree of silicate polymerisation increases, the solubility of silicate species in solution decreases, leading to higher viscosity, altered system charge balance and changes in setting behaviour of the synthesised geopolymer [20, 25, 26]. For these reasons, species with increasing polymerisation – especially fully polymerised colloidal silicas ($Q^4$) with (theoretically) no bridging oxygens – are generally considered less reactive and thus unfavourable within geopolymer activator solutions [21, 24]. Table 1 provides a graphical representation of several silica species of varying charges ($z$) and geometries (subscripts: *c*, *sq* for cyclic and square structures, respectively) in alkaline silicate (activator) solutions, adapted from Provis et al. [21] and Zhao et al. [27].

**Table 1:** Illustration of several silica species ($Q^n$) with varying charges ($z_i$) and geometries (subscripts: *c*, *sq*) in activator solutions from Provis et al. [21] and a simplified diagram of colloidal silica ($Q^4$) adapted from Zhao et al. [27].

| Charge ($z_i$) → | 0 | 1 | 2 | 3 |
|---|---|---|---|---|
| $Q^0$ | 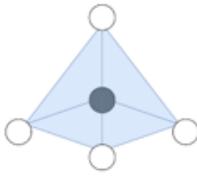 | 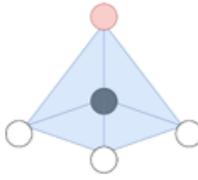 | 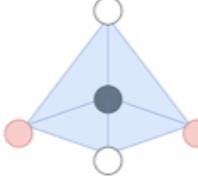 | 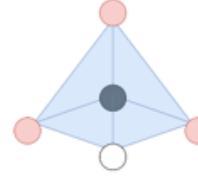 |
| $Q^1$ | 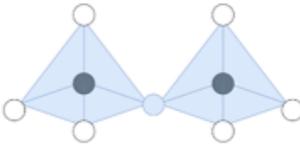 | 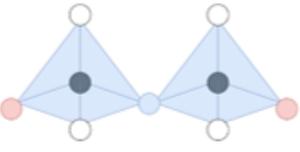 | 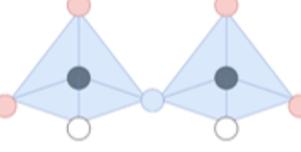 | 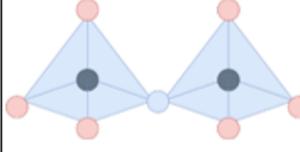 |
| $Q^2$ | 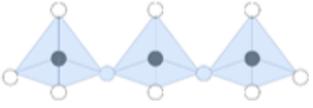 | 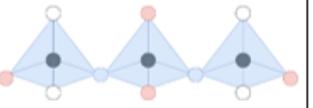 | 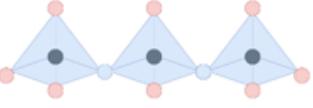 | - |
| $Q^{2c}$ | 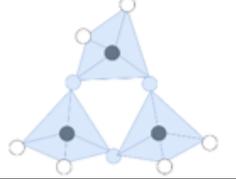 | 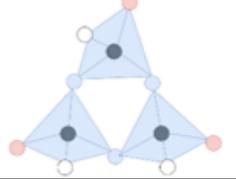 | 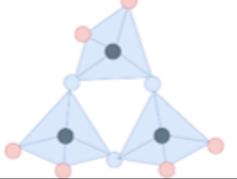 | - |
| $Q^3$ | 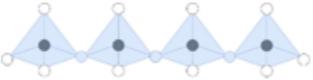 | 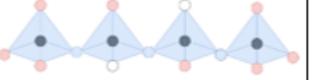 | - | - |
| $Q^{3c}$ | 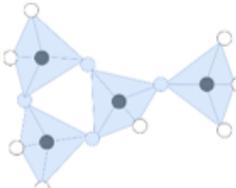 | 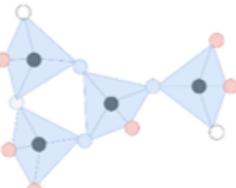 | - | - |
| $Q^{3sq}$ | 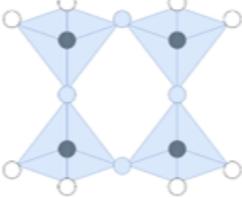 | 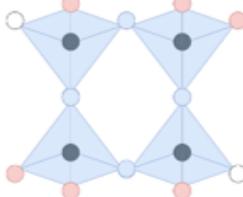 | - | - |
| $Q^4$ | \[Complex Charging\] 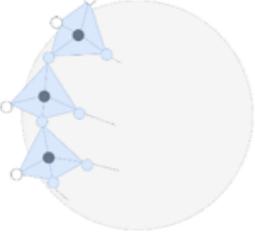 | | | |

Species ($Q^n$)

● = Si, ◯ = O, ○ = OH, ● = O⁻

Ternary systems of $H_2O$-$Na_2O$-$SiO_2$ in solution are classified according to their composition with solution boundaries dependent largely on their solubility, viscosity and commercial viability. This classification was determined by Vail [28] as displayed in the adapted ternary isotherm in Figure 2. This ternary diagram can be extrapolated for applications relevant to geopolymer activator solutions, with regions of interest labelled A, B, C and D. Each of these labels correspond to the contour area representing solutions classified as:

- (A): dilute aqueous solutions of NaOH and $SiO_2$.
- (B): "ordinary commercial [sodium silicate] liquids".
- (C): metastable "partially crystallised mixtures" of concentrated NaOH and $SiO_2$.
- (D): unstable "crystallised[/precipitated] alkaline silicates".

Most geopolymer activator solutions synthesised in the literature fall within region C in Figure 2. They are characterised by their partially solubilised "metastable" composition leading to inevitable precipitation over undefined periods which can span hours to days [18, 29]. Near or within region D, solutions typically precipitate rapidly into unstable gelatinous hydrated sodium metasilicates which are highly viscous and impractical to manually handle. In contrast, solutions with compositions in regions A and B represent "stable" solubilised analytes with lower viscosity, easier handling and are more predictable. For these reasons, commercial sodium silicates are usually only produced in region B, balancing technical stability, alkalinity and concentrated silica with commercial feasibility [28, 20]. Other unlabelled regions in Figure 2 represent compositions not commonly synthesised for geopolymer activators are not of interest in this study.

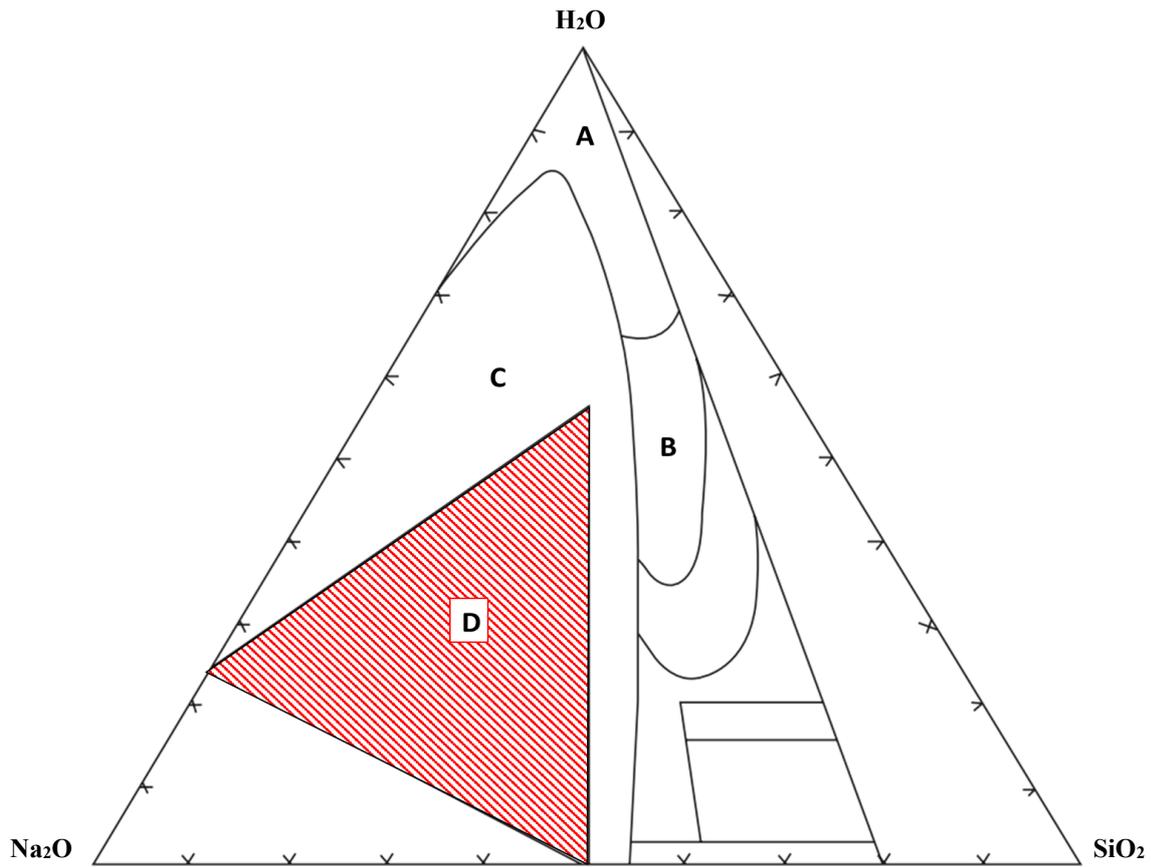

**Figure 2:** Ternary diagram of $H_2O$-$Na_2O$-$SiO_2$ system adapted from Vail [28]. Labels represent (A) dilute aqueous solutions of NaOH and $SiO_2$, (B) "ordinary commercial [sodium silicate] liquids", (C) "partially crystallised mixtures" of concentrated NaOH and $SiO_2$ (where most geopolymer activator solutions are synthesised) and (D) "crystallised[/precipitated] alkaline silicates".

## 1.2 Geopolymer Activator Solution Stability

Activator composition, solubility and silicate speciation are highly sensitive to synthesis conditions including temperature, order of mixing, feedstock purity, feedstock type, mix design and pH (among others). These variables can be used as predictors of activator solution stability and reproducibility. However, they are frequently underreported in the literature, complicating reproducibility even within the same laboratory where $SiO_2$ and NaOH solubilities vary with temperature, which may fluctuate seasonally in uncontrolled lab environments [16, 30]. Additional variability can arise from differences in feedstock mix sequences, inconsistent handling practices, or inadequate quality control of feedstocks [16].

Previous work (Skane et al. [16]) reviewed these inconsistencies and proposed a thermodynamic model to quantify activator solution stability. The model predicts the minimum period of enthalpic stability to be approximately one minute from initial mixing of the activator feedstocks – during which activator solutions are ready for combination with aluminosilicate precursors (e.g. fly ash, metakaolin). This contrasts with the extended and often undefined "equilibration" periods (i.e. ≥ 24 hours) reported in some literature. While this work offers a practical framework for determining when an activator is thermodynamically stabilised and 'ready' for geopolymerisation, it does not define the maximum stability period. This maximum period is defined by the time at which the solution becomes unstable due to reduced solubility from cooling and the onset of precipitation or gelation of the alkali-silicates. Quantification of both the minimum and maximum stability periods would enable the definition of a time-stability window for activator solutions. This window could be used as a predictive tool for process optimisation, quality control, and reproducible synthesis of geopolymers and their activators across varying feedstocks. Furthermore, the model does not incorporate critical factors such as chemical speciation, solubility limits, or silicate structures – parameters that strongly influence solution stability, reactivity and geopolymer performance (as mentioned in Section 1.1). These aspects form the focus of the present study alongside the quantification of activator time-stability windows.

## 2 Experimental Procedure & Mathematical Model

### 2.1 Feedstock Materials, Calorimetry & Thermodynamic Modelling

Experimental procedures for the calorimetry and preparation of activator solutions alongside feedstock sourcing, mathematical modelling and notation were identical to the work from Skane et al. [16]. The same dataset is provided for convenience in Table 2 and chosen to ensure that chemical speciation changes in solution arising from external temperature fluctuations were minimised. Modifications to the model are described in the subsections below with constants, user-specified values and equations entered and solved simultaneously using the Engineering Equation Solver software package (EES) [31].

**Table 2:** Activator Solution Design Compositions from Skane et al. [16].

| Experiment/ Solution ID | NaOH Solution $C_{NaOH}$ [M] | Final Activator Solution | | |
|---|---|---|---|---|
| | | $m_{SS}$ [g] | $C_{NaOH}$ [M] | Molar $SiO_2/Na_2O$ |
| A | 14 | 85.6 | 10.1 | 0.7 |
| B | 12 | 85.6 | 9.1 | 0.8 |
| C | 10 | 85.6 | 8.1 | 0.9 |
| D | 8 | 85.6 | 7.1 | 1.0 |
| E | 6 | 85.6 | 6.1 | 1.2 |
| F | 4 | 85.6 | 5.1 | 1.4 |
| G | 3 | 85.6 | 4.6 | 1.6 |
| H | 2 | 85.6 | 4.1 | 1.8 |
| SS* | - | - | 6.3 | 2.4 |

* Pure Sodium Silicate ("SS") Feedstock for reference

## 2.2 Nuclear Magnetic Resonance Spectroscopy

Quantitative Nuclear Magnetic Resonance Spectroscopy (qNMR) analysis of the $Q^n$ silica aggregates within the solvent activator solutions were conducted with $^{29}$Si NMR spectra acquired on a Bruker Avance III NanoBay 9.4 T spectrometer ($^{29}$Si, 79.486 MHz) equipped with a commercial 5 mm broadband observe probe and 1,280 transients acquired using a single 90° pulse of 18.8 µs, relaxation delay of 5 s, and acquisition time of 1.02 s. The NMR spectra were processed using Bruker's TOPSPIN program to quantify the individualised silica species (i.e. $Q^0$, $Q^1$, $Q^{2c}$, $Q^2$, $Q^{3c}$, $Q^3$ and $Q^4$) of the sodium silicate and activator solutions at steady state, all of which were referenced to monomeric silicate.

## 2.3 Chemical Speciation Modelling

NMR data was used to quantify activator solution silica speciation and modelled as a function of its $SiO_2/Na_2O$ molar ratio. This ratio was compared with established literature findings from Provis et al. [21] and Harris et al. [32] between molar ranges of $0.25 \leq SiO_2/Na_2O \leq 2$ and $2 < SiO_2/Na_2O \leq 4$, respectively. Both models were merged into a single hybridised map (Figure C.1, Appendix C) spanning a wide $SiO_2/Na_2O$ range to include the sodium silicate feedstock (i.e. $SiO_2/Na_2O = 2.41$) and activator solutions (typically with a $SiO_2/Na_2O \leq 2$) used in this study. While combining both models introduces a discontinuity in $Q^n$ species at $SiO_2/Na_2O = 2$, leading to uncertainty for experimental values around this ratio, this approach was preferred to better account for the quantification of colloidal silica ($Q^4$) aggregates and the dynamic transformations of all species. The fine siliceous complexes characteristic of $Q^4$ colloidal silica, more prevalent at higher molar ratios ($SiO_2/Na_2O > 2$), are known to exist in solution but are only considered in the Harris et al. model [32]. For simplicity, the combined model summarises all species of different geometries (Table 1) into a total dissolved fraction ($Q^n_{Tot}$, e.g. $Q^2_{Tot}$ [wt.%] = $Q^2 + Q^{2c}$ etc.), excluding colloidal silica $Q^4$, at $T_{ref}$.

These total dissolved species were used as the initial conditions ($Q_{Tot,i}^n$) in the steady-state model (Equation 1) to estimate the speciation of activator solutions at a given temperature, $T_{Soln}$ (i.e. under non-isothermal conditions). Temperature-dependent speciation trends were derived from Kinrade and Swaddle [33] (conveyed in Figure C.2, Appendix C) and reformulating into differentiable form ($\frac{\partial Q_{Tot}^n}{\partial T}\Delta T$ [wt.%]) via one-sided backwards finite difference methods (expanded in Appendix B). This allowed the extrapolation of each species according to Equation 1 with notation as per Skane et al. [16].

$$Q_T^n(Q_{Tot,i}^n, T_{soln}) = \omega(t)_{SS}\left(Q_{Tot,i}^n + \frac{\partial Q_{Tot}^n}{\partial T}\cdot(T_{Soln} - T_{ref})\right) \qquad (1)$$

## 2.4 Solubility Modelling

The complexity of alkali-silicate activator solution systems arises from the interdependent solubilities of concentrated NaOH and SiO$_2$, along with their hydrated species formed upon moisture exposure. In Figure 2, these solubility limits are visualised as region-partitioning contours that separate stable, metastable and unstable solution domains. Any point on the ternary diagram, including the contour solubility boundaries, are mapped as a function of the solution's H$_2$O wt.%, Na$_2$O wt.% and SiO$_2$ wt.% at an isothermal (reference) temperature ($T_{ref}$ = 25°C) [28]. These compositions can be expressed more simply as the solution's concentrated NaOH, $C_{NaOH}$, and SiO$_2$, $C_{SiO_2}$. The first solubility boundary of interest in Figure 2 is visualised as the 'triangular' contour partitioning regions C and D, with unstable precipitated alkaline silicates existing in region D. This trajectory represents the border of metastable activators and highly viscous unstable ones. The second boundary partitions metastable solutions in region C from all other regions, including stable solutions in regions A and B. To quantify activator solubility boundaries, each of these contours were discretised along their trajectories into oxide weight fractions and converted into molar $C_{NaOH}$ and $C_{SiO_2}$ using solution density calculations from literature and $T_{ref}$ from EES' in-built multivariable density functions [34]. The resulting {$C_{SiO2}, C_{NaOH}$} dataset was then approximated by a linear regression fit to express each metastable and unstable ("*MS*" and "*US*", respectively) boundary as polynomial functions. For convenience, the *MS* and *US* solubility limits are expressed as elements of the set $X$ in the equations below (i.e. $X \in \{MS, US\}$ ). In this way, each solubility boundary is formalised as a bivariate map as per Equation 2 whereby a given silica concentration and temperature returns the NaOH solubility concentration that remains on the $X$ boundary. Note that the temperature range $T \in [15,95]$ is discussed further in this section.

$$C_X : \{C_{SiO2} \in [0,8] \ (mol/L)\} \times \{T \in [15,95] \ (°C)\} \to C_{NaOH}(mol/L), \quad X \in \{MS, US\} \quad (2)$$

The unstable boundary, $C_{US}(C_{SiO2}; T_{ref})$, is expressed in the piecewise Equation 3 (rows 1-2) which was approximated by a two-segment linear regression with its y-intercept representing the isothermal saturation of $C_{NaOH}$ in the H$_2$O-Na$_2$O system. The hexic metastable polynomial approximation, $C_{MS}(C_{SiO2}; T_{ref})$ ($R^2$ = 99.54) is presented in Appendix B for brevity but shown as row 3 in Equation 3. This polynomial fit, along with all others used in this work, with coefficients $a_k$ and corresponding degree $n_d$ as per the general Equation 4 are provided in Appendix B for conciseness.

$$C_X(C_{SiO2}; T_{ref}) = \begin{cases} 24.13 - 4.92 C_{SiO2}, & X = US, C_{SiO2} \leq 3.47 \\ 0.33 + 1.95 C_{SiO2}, & X = US, C_{SiO2} > 3.47 \\ [(Appendix\ B)\ poly.\ in\ C_{SiO2}], & X = MS \end{cases} \quad (3)$$

$$\sum_{k=0}^{n_d} a_k x^k \quad (4)$$

To capture the continuous temperature solubility dependence of NaOH(aq) and SiO$_2$ (quartz), solubility curves were sourced from literature and used to evaluate partial derivative expressions (i.e. $\frac{\partial C_X}{\partial T}$). For NaOH, solubility data were compiled over a 0-200 °C range and fitted by least-squares to a hexic polynomial ($R^2$ = 99.33) [35, 36]. This function was constrained over a practical domain of interest between 15-95 °C to reflect activator solutions in literature and the experimental data, whilst also mitigating high-degree oscillations (Runge phenomenon) with general fitting of the narrower range. These oscillations appear due to the nonlinear curvature of the dataset, an artefact of complex NaOH hydrate precipitates affecting overall solubility dynamics [35]. The same approach was used with SiO$_2$ solubility data with an identical temperature constraint used for consistency, however, this constraint is technically unnecessary due to the smoother dataset [30]. Both functions are presented in Appendix B.

The above expressions were then integrated together to formulate hypersurface equations defining both metastable and unstable limits of activator solutions as functions of the solubility limits and a given solution temperature in Equation 5. Typically, this solution temperature, $T_{Soln}$, is the minimised stability temperature at which activator solutions become thermodynamically stable after addition of all activator feedstocks, as quantified from Skane et al. [16]. Equation 6 provides an extended term that can be added to Equation 5 for dynamic simulation of solubility changes from a given $T_{Soln}$ as it changes to any other actual solution temperature (e.g. cooling from the mixed and initially stable temperature $T_{Soln}$ to the ambient temperature $T_{Amb}$).

$$C_X(C_{SiO2}; T_{Soln}) = C_X(C_{SiO2}; T_{ref}) + \left.\frac{\partial C_X}{\partial T}\right|_{T_{ref}} (T_{Soln} - T_{ref}) \qquad (5)$$

$$C_X(C_{SiO2}; T_{Amb}) = C_X(C_{SiO2}; T_{Soln}) + \left.\frac{\partial C_X}{\partial T}\right|_{T_{Soln}} (T_{Amb} - T_{Soln}) \qquad (6)$$

Based on these definitions and by equation rearrangement, any activator solution system – characterised by its SiO$_2$ and NaOH concentrations and temperature (i.e. $(x, y, z) = (C_{SiO2,f}, C_{NaOH,f}, T_{Soln})$ – can be mapped in three-dimensional space relative to its temperature dependent metastable and unstable solubility limits in Equation 7 (where $C_{NaOH,f}$ is contextualised in row 2 via Equation 3). In this equation, $\lambda_X$ is a scalar multiplier of the "stability vector" (the $X$ hypersurface normal) which measures how far and in which direction the given solution state must move to intersect the *MS* or *US* solubility surface.

$$\text{For each } X \in \{MS, US\}: f_X\left(C_{SiO2,f}, C_{NaOH_f}, T_{Soln}\right) = \begin{cases} C_{SiO2,i} = C_{SiO2,f} + \lambda_X \left.\frac{\partial C_X}{\partial C_{SiO2}}\right|_{C_{SiO2,f}}, \\ C_{NaOH,i} = C_X(C_{SiO2}; T_{Soln}) - \lambda_X, \\ T_i = T_{Soln} + \lambda_X \left.\frac{\partial C_X}{\partial T}\right|_{T_{Soln}} \end{cases} \qquad (7_1)$$

This thermophysical relationship can be extended further by quantifying the minimised distance between the given ($x, y, z$) point to the nearest non-stable hypersurface. In this context, an activator solution's non-stability temperatures, $T_{\lambda MS}$ and $T_{\lambda US}$, can be defined by finding the roots of $\lambda_{MS}(T) = 0$ and $\lambda_{US}(T) = 0$ (summarised in Table 3). These zero-points correspond to the temperatures at which a cooling activator crosses into the metastable or unstable solution space, and at an increasing degree of precipitation proportional to the negative magnitude of $\lambda_X$.

**Table 3:** Activator Solution Stability State Definitions and Temperature Milestones.

| Solution State | Solubility Concentration Exceeded | Stability Conditions | | Temperature Milestones |
|---|---|---|---|---|
| | | $\lambda_{MS}$ | $\lambda_{US}$ | |
| "Stable" | None | + | + | $T_{Soln} > T_{\lambda MS} > T_{\lambda US}$ |
| "Metastable" | $C_{SiO2}$ | ≤ 0 | + | $T_{\lambda MS} \geq T_{Soln} > T_{\lambda US}$ |
| "Unstable" | $C_{SiO2}$ & $C_{NaOH}$ | ≤ 0 | ≤ 0 | $T_{\lambda MS} > T_{\lambda US} \geq T_{Soln}$ |

## 3   Results & Discussion

### 3.1   qNMR Analysis

[29]Si NMR spectra of the Experimental C, D and E activator solutions, alongside the sodium silicate parent feedstock are provided in Figure 3. Spectral analyses for solutions A and B were discontinued prior to measurement due to overnight precipitation into gelatinous solids, which prevented clear data acquisition using the NMR instrument in this study. Although over 20 different silica oligomers have been identified in literature using [29]Si NMR [22], this study focuses only on the integrated peaks ascribed to the known $Q^n$ and $Q_c^n$ silica species at similar chemical shifts to those reported in previous research. The subscript $c$ distinguishes $Q_c^2$ and $Q_c^3$ silica sites present as part of cyclic rings from their acyclic counterparts ($Q^2$ and $Q^3$), which appear at different resonances and downfield peak shifts. These cyclic species occur throughout the spectra due to deshielding effects from their different geometries, despite having the same number of silicon atoms. Quantified NMR spectral peak shifts, speciation and integrated areas are summarised in Table 4 which align with literature for this SiO$_2$/Na$_2$O range [21, 33, 24, 26, 37] and are in agreement with the combined model previously presented [21, 32].

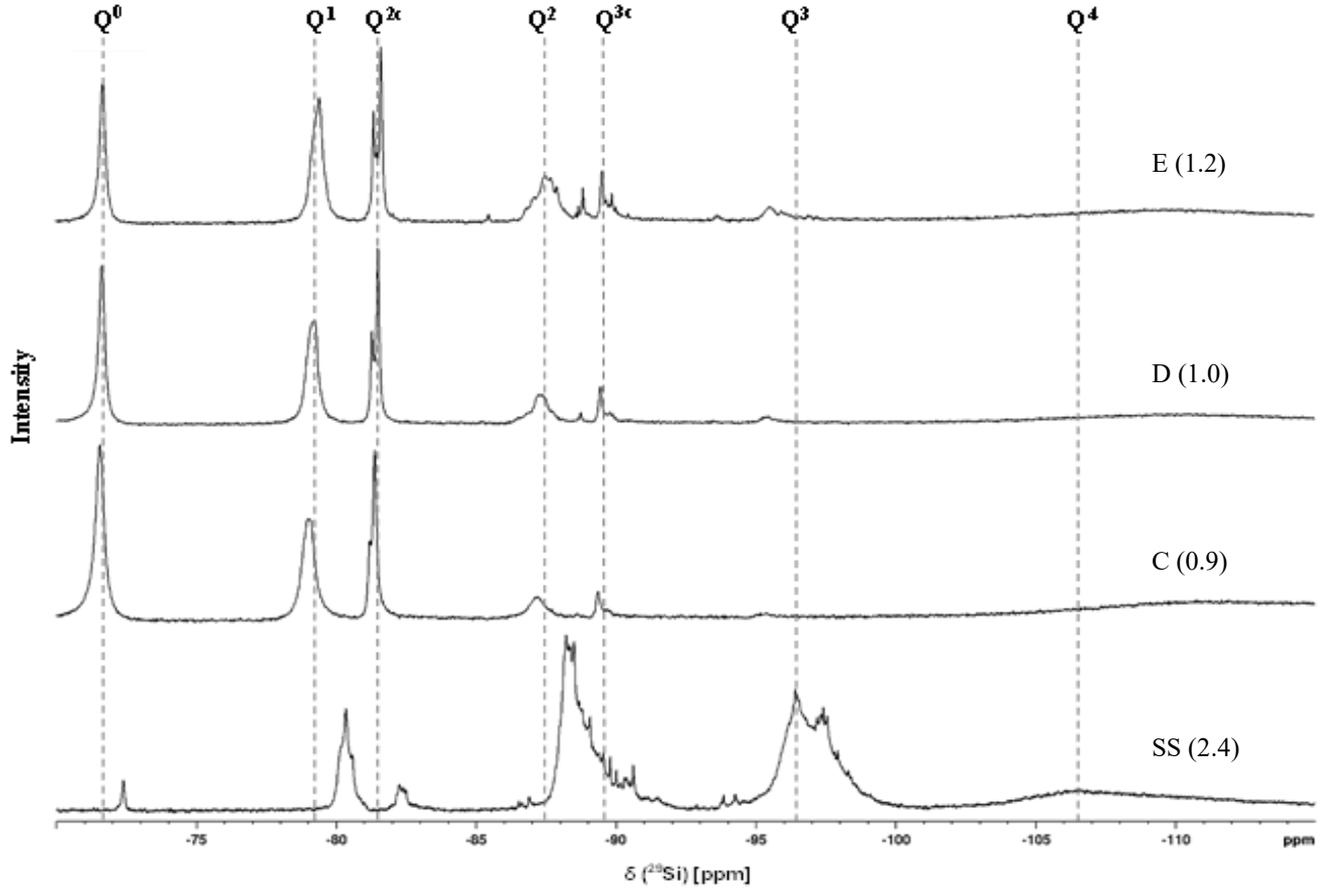

**Figure 3**: $^{29}$Si NMR spectra of experimental activator solutions labelled with their characteristic SiO$_2$/Na$_2$O molar ratios and $Q^n$ and $Q_c^n$ spectra outlined.

**Table 4:** Quantified chemical speciation from $^{29}$Si NMR spectra of experimental activator solutions.

|  |  |  | Activator Solution | | | |
|---|---|---|---|---|---|---|
|  |  | Sample ID | C* | D | E | SS |
|  |  | Final C$_{NaOH}$ [M] | 8.1 | 7.1 | 6.1 | 6.3 |
|  |  | Molar SiO$_2$/Na$_2$O | 0.9 | 1.0 | 1.2 | 2.4 |
| Silicate Species | $Q^0$ | δ [ppm] | -71.6 | -71.6 | -71.7 | -72.4 |
|  |  | Area [%] | 38.4% | 31.7% | 19.4% | 1.0% |
|  | $Q^1$ | δ [ppm] | -79.0 | -79.2 | -79.4 | -80.3 |
|  |  | Area [%] | 28.2% | 31.3% | 27.0% | 9.1% |
|  | $Q^{2c}$ | δ [ppm] | -81.4 | -81.5 | -81.6 | -82.3 |
|  |  | Area [%] | 19.1% | 15.8% | 19.2% | 2.7% |
|  | $Q^2$ | δ [ppm] | -87.2 | -87.2 | -87.5 | -88.2 |
|  |  | Area [%] | 8.4% | 13.0% | 17.7% | 36.9% |
|  | $Q^{3c}$ | δ [ppm] | -89.4 | -89.5 | -89.6 | -90.6 |
|  |  | Area [%] | 4.3% | 5.6% | 10.7% | 1.0% |

|   |         |        |        |        |        |
|---|---------|--------|--------|--------|--------|
| $Q^3$ | δ [ppm]   | -95.4  | -95.5  | -95.5  | -96.4  |
|   | Area [%]  | 1.6%   | 2.5%   | 6.0%   | 42.3%  |
| $Q^4$ | δ [ppm]   | -111.7 | -110.2 | -109.9 | -106.6 |
|   | Area [%]  | 0.0%   | 0.0%   | 0.0%   | 7.0%   |

* Liquid component of solution was analysed after observing solid precipitates form, details in Appendix C.

## 3.2 Thermochemical Stability

### 3.2.1 Chemical Speciation Modelling

The combined speciation model from Section 2.3 is displayed in Figure 4 where each species is represented as a surface dependent on both the molar $SiO_2/Na_2O$ and solution temperature. For clarity, cyclic and acyclic species were summed into $Q^2_{Tot}$ and $Q^3_{Tot}$ variables, while $Q^4$ silica (colloidal, not dissolved) was omitted due to the lack of temperature-dependent speciation data in the Kinrade and Swaddle model [33]. Whilst the speciation surfaces smoothly depict silica dynamics in solution, a noticeable transition occurs for all species at $SiO_2/Na_2O$ = 2, where both literature and model $SiO_2/Na_2O$ boundaries diverge. While interpolation could improve accuracy at this boundary, it was deemed unnecessary for this study as the examined activator solutions primarily fall within $SiO_2/Na_2O$ < 2. However, the full plot is provided for general reference, particularly for laboratory investigations of solutions with $SiO_2/Na_2O$ > 2.

Overall, the adapted model predicts an increase in oligomeric silica speciation with increasing molar $SiO_2/Na_2O$ in the activator solutions. In general, at a given molar ratio where $n$ < 4 species are present, higher temperatures promote proportionately depolymerised species formation at the expense of $Q^3_{Tot}$. Since lower-polymerised species contain a higher number of non-binding oxygen (NBO) sites available for reaction, these results suggest that higher temperature reduces silica polymerisation and can improve the solution's theoretical reactivity [24, 26]. This finding may be advantageous for activator solutions paired with other geopolymer precursors with a low tendency to react, cement or otherwise have aluminosilicate materials that are unreactive at lower ambient temperatures. Conversely, excessive silica depolymerisation may lead to geopolymer gels that react too quickly (i.e. "flash set"), in which case the activator solution should be cooled (or allowed to lose heat) before mixing. Prediction of specific activator solution temperatures to optimise these processes is presented in Figure 5 for the Experiment E solution at given steady state temperatures which is shown as an example for how the optimisable activator system can be monitored. The combined equation set from the models provided by Skane et al. [16]

can be used to model dynamic speciation in activator solutions and would likely be complementary with the speciation findings of these results. However, such an extended model would need to accommodate for complex chemical speciation kinetics to ensure accurate transient analyses. Literature relating to silica speciation kinetics shows dynamic changes in all $Q^n$ reaching chemical equilibration within a matter of seconds, or more conservatively, over a 1–2-minute period in proportion to the degree of (non-monomeric) polymerisation [38, 39]. Thermodynamically, these chemical changes have been associated with the enthalpy of mixing of sodium silicate [40], however, its investigation is considered out of scope for this work.

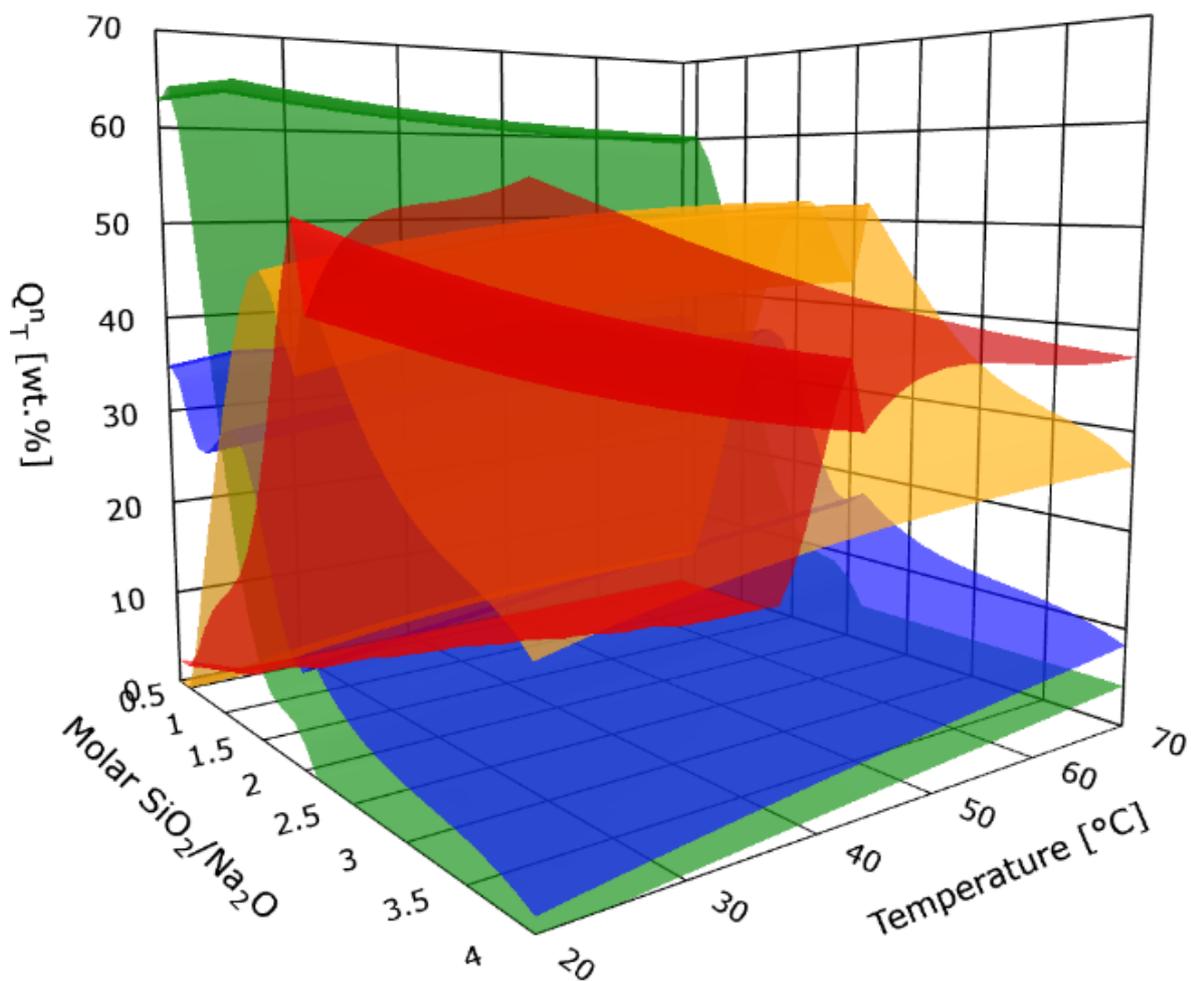

**Figure 4**: Model calculations for the solution space conveying activator solution speciation with varying temperatures where the coloured surfaces represent $Q^0$ (green), $Q^1$ (blue), $Q^2_{Tot}$ (yellow) and $Q^3_{Tot}$ (red).

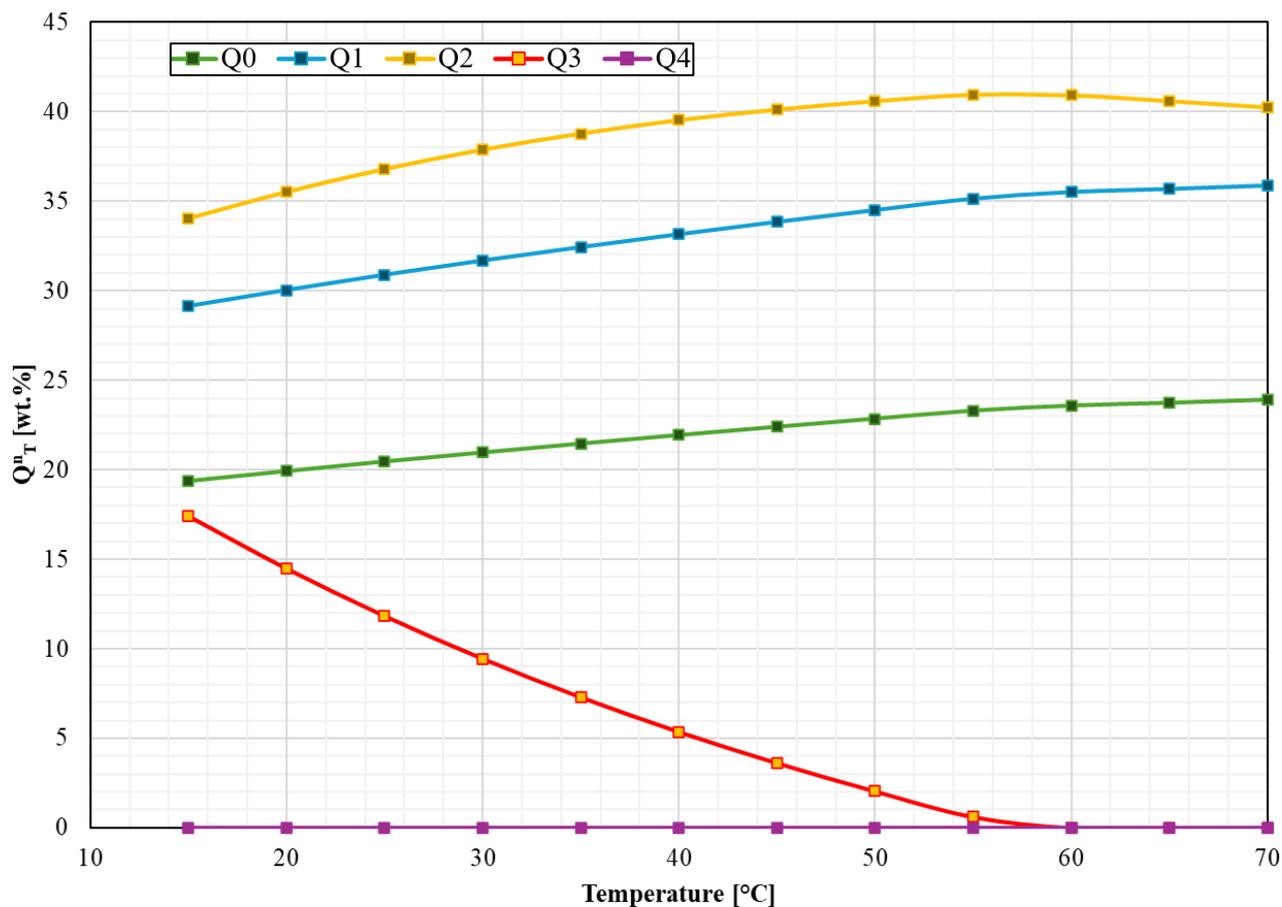

**Figure 5**: Steady state modelling results for Experiment E solution conveying activator solution speciation with varying temperatures where the coloured trends represent $Q^0$ (green), $Q^1$ (blue), $Q^2_{Tot}$ (yellow) and $Q^3_{Tot}$ (red).

Before running the Experiment C solution in the NMR, solid precipitates were qualitatively observed adhering to the walls of the measuring vial, perhaps crystallising along a temperature gradient in the solution. For this unstable solution, a decant of the liquid component was analysed. It was initially anticipated that the resulting NMR spectrum would deviate significantly from comparable literature data, as silica previously solubilised as $Q^n$ species would have been partially removed through precipitation. However, the results for this experiment quantified in Table 4 and spectra visualised in Figure 3 show a surprisingly similar agreement with literature findings (Appendix C, Figure C.1). These results suggest that either (1) the precipitated and dissolved species were compositionally similar (though unlikely given their differing solubilities), or (2) that precipitation occurred primarily after cooling to temperatures lower (presumably due to the general decreases in NaOH and $SiO_2$ solubility with temperature [35, 30]) than the 25°C isothermal reference used in the model [21]. As all solutions experienced overnight temperatures before analysis with lower laboratory temperatures reaching 15-20°C, the latter case is supported, alongside the chemical stability findings reported later in Section 3.2.2.

These observations underscore an important limitation in that conventional liquid-phase NMR techniques capture only the solubilised silica species and omit any metastable or unstable solids. By not accounting for the precipitated silica, they may not fully reflect the optimised reactivity or "effective silica" available for geopolymerisation. It is unknown what effect these silica precipitates have when added to a geopolymer precursor or what impacts, if any, its absent measurement has throughout the reported synthesised geopolymer properties. At worst, their unreactive state is postulated to inhibit geopolymerisation, preclude efforts for quality control and potentially become deleterious to the material's durability; at best, their presence in the activator solution suggests an inefficient mix design in which soluble silica feedstocks (usually the most expensive feedstock in geopolymer systems [6]) are unnecessarily wasted.

### 3.2.2 Solubility

The activator thermochemical solution space proposed by the mathematical model in Section 2.4, is presented in Figure 6 and represents a different activator solution at each 3-dimensional point as defined by its solution temperature and concentrated $SiO_2$ and NaOH. This space is populated by two characteristic hypersurfaces which each represent the metastable and unstable solubility boundaries of a given geopolymer activator solution from Equation 5 (i.e. $X = US$). Due to the low solubility of quartz (i.e. ~10 ppm at 25 °C), its hypersurface has been omitted from Figure 6 for clarity, but is provided in Appendix D. Whilst stable alkali-silicate solutions can exist in regions A and B from Figure 2, practically, most activator solutions in literature (and all as mentioned as part of this work) are synthesised in region C as based on the desired Si/Al and Na/Al molar ratios of typical feedstocks and geopolymer systems. Following this, the results of this work and its extrapolations only assumes solutions that have been synthesised in region C and are, at a minimum, considered metastable.

The inclusion of the unstable surface partitions the solution space into 2 distinct regions (volumes) of stability in which activator solutions exist transiently as they cool. The first region, above the surface, is the metastable region in which all simulated activator solutions are relatively soluble and are observed to exist at temperatures higher than 50°C. The second region below the surface defines activator solutions which are chemically unstable with formed precipitates of $SiO_2$ and NaOH combining to form heterogenous and highly viscous gels which are not recommended for synthesis. Such solutions consist of increasing thermochemical heterogeneity, very poor repeatability, physical handling difficulties and extremely corrosive NaOH safety concentrations. The solution space presented at 25°C represents good agreement with the isotherm ternary plot in Figure 2, shown previously, however, there is a slight discrepancy between the ternary plot's extrapolated NaOH solubility point of 24.1 mol/L (i.e. the y-intercept in Equation 3) [28] and reported literature values of 25 mol/L [36]. This disparity is considered negligible for the purposes of this work resulting from the propagated error arising from the extrapolative methods (outlined in Section 2.4) of the manually drawn contours in Figure 2.

To demonstrate its practicality, the A, C and E experimental solutions are provided as squares, diamonds and circles. Orange markers represent the initial post-sodium-silicate stabilised mixing temperatures (i.e. $T_{SS,f}$, relatively higher for experiments with higher NaOH concentrations [16]), while blue markers represent their eventual ambient temperatures (25 °C). Across all cases, the vertical descent along the z-axis (temperature) corresponds to heat loss over time. During this cooling process, both the experimental A and C solutions have descended past the unstable surface boundary whilst the E solution remains just above the surface (with their minor differences due to the unstable surface's curvature). Complementing these results, both A and C solutions were qualitatively observed to precipitate in the laboratory with the former explained previously (i.e. difficulty in NMR and chemical speciation testing) and the latter forming a thick white and unusable gel. Solution E, while still viscous, remained translucent and above the unstable boundary. These transitions confirm the model's ability to predict instability as a function of cooling trajectory.

Regardless of the counterproductive quality control considerations of solutions in the unstable or highly metastable regions, the increasingly (i.e. as solutions continue cooling/traversing towards instability) high viscosity of the solids formed in regions of instability or high metastability are practically incapable of manipulation without the use of sharp scraping tools and/or (uneconomical) external heating. While reheating can move a solution back to a higher temperature region (along the z-axis) with a more theoretically favourable stability, practically, the unstable phase transition is considered irreversible due to thermochemical hysteresis – a phenomenon linked to the precipitated solution's rapid change in viscosity [41].

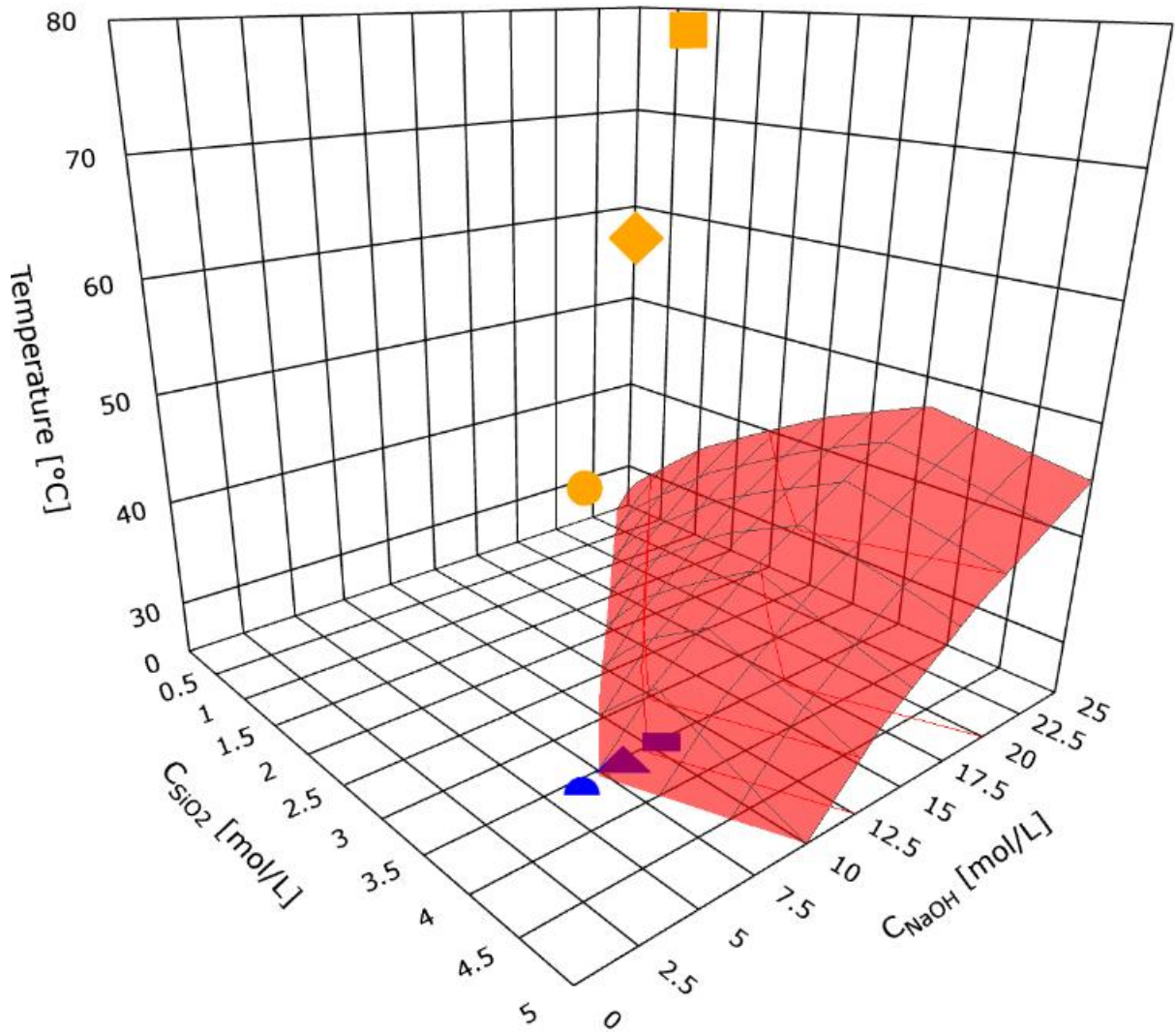

**Figure 6**: Activator solution stability solution space conveying metastable and unstable regions partitioned by the unstable ("US", red) hypersurface. Centred dot points representing the A, C and E experimental solutions plotted as squares, diamonds and circles where the orange and blue coloured icons represent the sodium silicate stabilisation (i.e. $T_{Stable}$) and ambient 25°C temperatures, respectively.

For practical purposes in avoiding transitioning to an unstable state, all solutions should be used promptly after the addition of sodium silicate whilst still at elevated temperatures from in-situ residual heat. This process also ensures thermodynamic and chemical speciation stability (i.e. a very low differential). This finding is in stark contrast with literature precedence (as summarised by Skane et al. [16]) and provides further supporting evidence against overnight or extended periods of mixing where activator solutions could conceivably cool down towards increasing non-stability / precipitation. Results in Table 5 and Figure 7 present the combined use of this study's solubility model and the thermodynamic cooling profiles from Skane et al. [16] in enabling the quantification of a 'time-stability window' bounded by two critical values: the minimum required stabilisation time and temperature ($t_{Stable}, T_{Stable}$), and the predicted point of thermochemical instability ($t_{\lambda US}, T_{\lambda US}$). As an example, Experiment C is shown to become unstable at 26.7 °C after 7.3 hours where, practically, the solution becomes a highly viscous and unusable gel. As such, this solution should not be allowed to cool to this critical point for maintaining solution stability. In contrast, experiments D and E stay in their originally formed metastable state (not intersecting with the unstable hypersurface in Figure 6 at ambient conditions).

**Table 5:** Stability temperatures for experimental alkaline solutions, to 1 dp.

| Experiment/ Solution ID | Molar $SiO_2/Na_2O$ | $H_2O$ [wt.%] | Stability Metrics (This Study, Calculated) | | | | Literature [d] | | |
|---|---|---|---|---|---|---|---|---|---|
| | | | $T_{Stable}$ [°C] [a] | $t_{Stable}$ [mins] [b] | $T_{\lambda US}$ [°C] | $t_{\lambda US}$ [hours] [c] | $T_{Stable} = T_{\lambda US}$ [°C] | $t_{Stable}$ [mins] | $t_{\lambda US}$ [hours] |
| A | 0.7 | 55% | 80.4 | 1.5 | 30.7 | 6.0 | N/A | ≥ 24 | N/A |
| B | 0.8 | 57% | 72.6 | 1.1 | 28.8 | 6.8 | N/A | ≥ 24 | N/A |
| C | 0.9 | 59% | 63.4 | 1.4 | 26.7 | 7.3 | N/A | ≥ 24 | N/A |
| D | 1.0 | 62% | 59.6 | 1.1 | 24.2 | *Metastable* | N/A | ≥ 24 | N/A |
| E | 1.2 | 64% | 50.5 | 1.3 | 21.5 | *Metastable* | N/A | ≥ 24 | N/A |

[a] Extrapolated initial stability temperatures $T_{Soln}(t_{Stable}) = T_{Stable}$ from the calculations behind Figure 7 in Skane et al. [16].

[b] Results from the calculations in Skane et al. [16].

[c] Extrapolated instability time $T_{Soln}(t) = T_{\lambda US}$ from Figure 7 in Skane et al. [16].

[d] Literature "stability" times (or points where the activator solution is used at mixing) which, when specified, can range from 1 – 24 hours. However, most times are unspecified or inconsistent (e.g. "mixed overnight") [16].

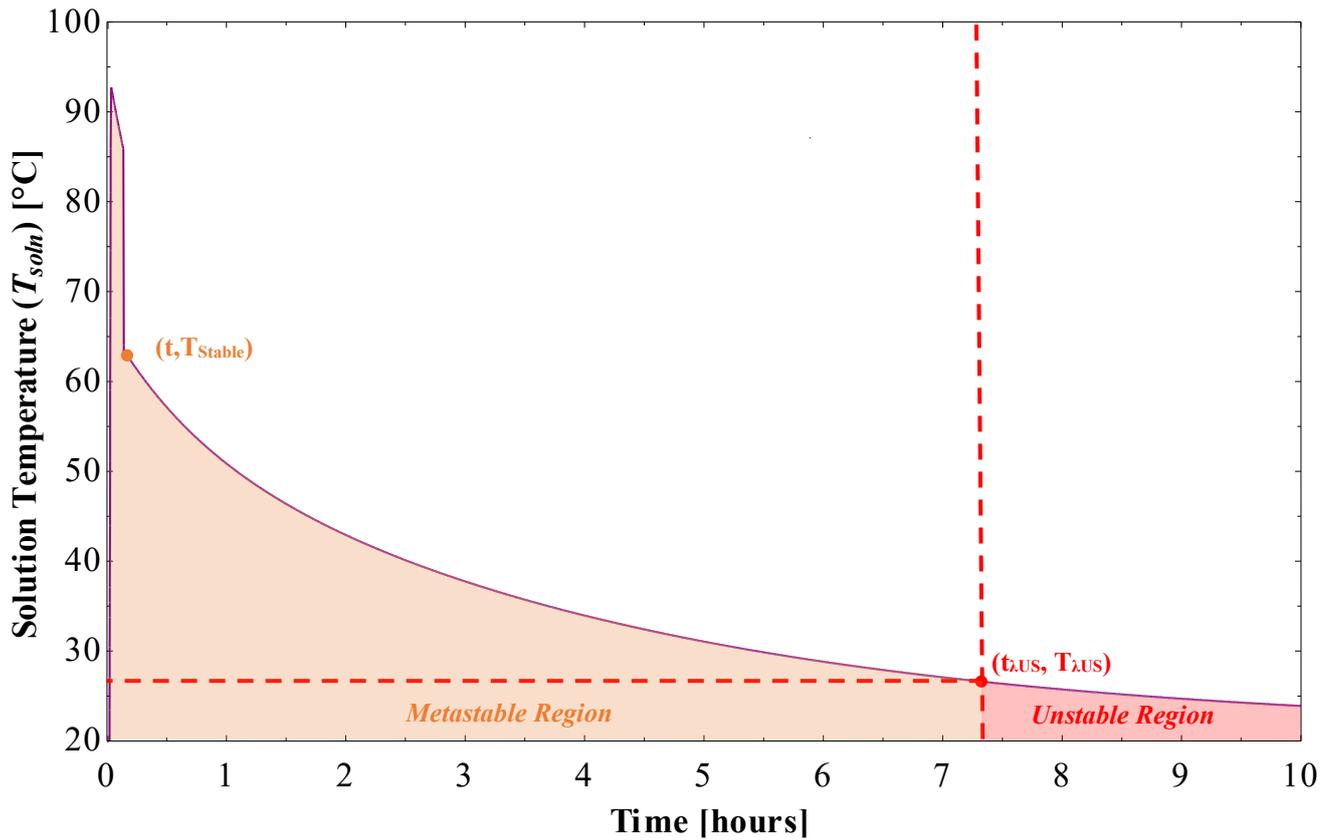

**Figure 7**: Modelled cooling time data from Skane et al. [16] of activator solution C with labelled stability milestones and regions.

Whilst seemingly counterintuitive, solutions presented in Table 5 with higher molar $SiO_2/Na_2O$ ratios (e.g., solution E) exhibit lower stability temperatures than those with lower ratios (e.g., solution A). This is due to the influence of dilution and temperature in the full ternary solution space. Molar $SiO_2/Na_2O$ is a simplified two-dimensional projection of a multi-dimensional space defined, in this paper at three-dimensions, by $Na_2O$, $SiO_2$, and $H_2O$. At fixed $SiO_2/Na_2O$ ratios, varying dilution can move a solution from a stable to unstable region (or vice versa). Therefore, apparent trends based on $SiO_2/Na_2O$ alone do not hold unless dilution is fixed – which it is not in these experiments due to practical difficulties in mix designing conventionally used activator solutions from literature. As shown in Table 5, increased $H_2O$ content correlates with lower $T_{Stable}$ values, reflecting the compound effects of dilution cooling from sensible heat transfer and temperature-dependent solubility. This highlights the importance of using concentration-based metrics (rather than molar ratios alone) when assessing stability in activator solutions.

With literature outlining speciation equilibration in a matter of seconds and thermodynamic stabilisation in a matter of minutes, these combined results support that activator solutions can be ready-for-use in a matter of minutes [16, 38, 39]. Furthermore, the extensive activator cooling or "equilibration" periods are regarded as technically unnecessary, time-inconvenient and in agreement with further literature that suggests the performance limitation of geopolymer systems made from cooler activators (i.e. which inhibit aluminosilicate precursor dissolution potential) [42]. The quantification of this window for a given activator solution can be used as a predictive tool for process optimisation, quality control, and reproducible synthesis of geopolymers and their activators across varying feedstocks.

### 3.3 Activator Feedstock Sequencing & Mixing

In extension to the dynamics explored as part of this study, sequencing of activator feedstock addition plays a critical role in optimising thermochemical stability and preventing transitions into non-stable regions. As shown in Figure 8, the order in which water, sodium hydroxide, and sodium silicate are combined directly affects whether the solution crosses solubility hypersurfaces that lead to undesirable precipitation.

The optimal mixing strategy, supported by this study, begins with mixing the binary $H_2O$-NaOH analyte system. After allowing approximately 1–1.5 minutes for thermodynamic stabilisation (see Table 5), the silicate feedstock is introduced to form the full ternary $H_2O$-NaOH-$SiO_2$ system. This sequence is visualised by the green arrows in Figure 8.1, which trace a stable path through Regions A and C of the isotherm. Instability arises only if the solution cools excessively (i.e $T \leq T_{\lambda US}$). In contrast, reversing this order – by adding sodium silicate first, followed by NaOH and water—can push the system directly into the unstable Region D. This is represented by the red arrow in Figure 8.1, which traverses Region B → D → C. Due to thermochemical hysteresis, solutions that even briefly enter Region D may not be restored to stability even with reheating, and irreversible precipitation may occur. Note that the increased traversing into Region D proportionately increases the likelihood of instability due to the hypersurface's curvature in Figure 6, this can be minimised by adding feedstocks at times of thermodynamic stability [16].

These risks are demonstrated in Figure 8.2 (from Skane et al. [16]), where two activator solutions of identical composition (14.9 wt.% $SiO_2$, 20.3 wt.% $Na_2O$, 64.8 wt.% $H_2O$) display vastly different outcomes. The left-hand solution, prepared with informed sequencing, remains stable and fluid. The right-hand solution, improperly sequenced, forms a gelatinous precipitate, becoming unusable. These outcomes occurred under identical ambient conditions and illustrate how feedstock order – and not just chemical composition – is an additional defining variable which can be used towards optimisation. Practitioners unaware of this process control risk discarding or using unstable solutions with poor repeatability, despite the availability of a simple and effective preparation method using the same feedstocks with a different addition sequence.

In this way, the results provided as part of this research provides further evidence that does not support many of the methods referred to in the geopolymer literature when synthesising the geopolymer activator solution (as summarised by Skane et al. [16]) in which extensive and undefined periods of unqualified "equilibration" of activator solutions are conducted.

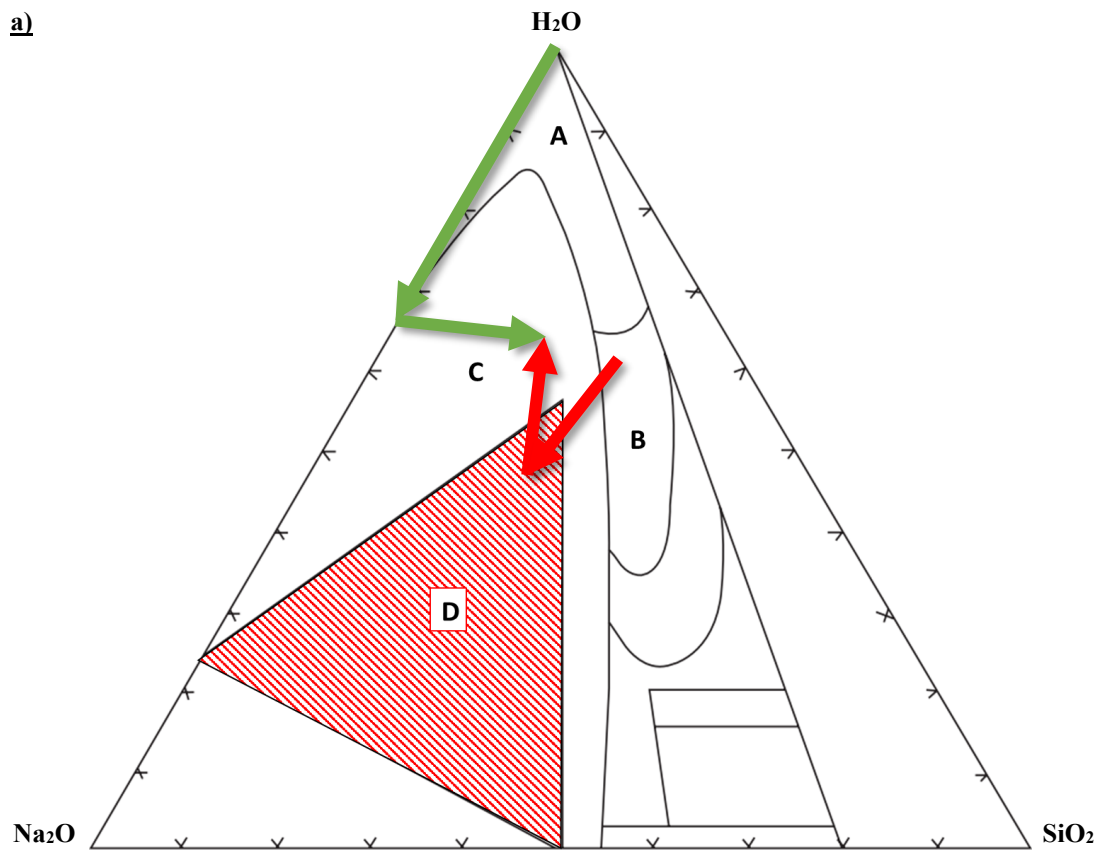

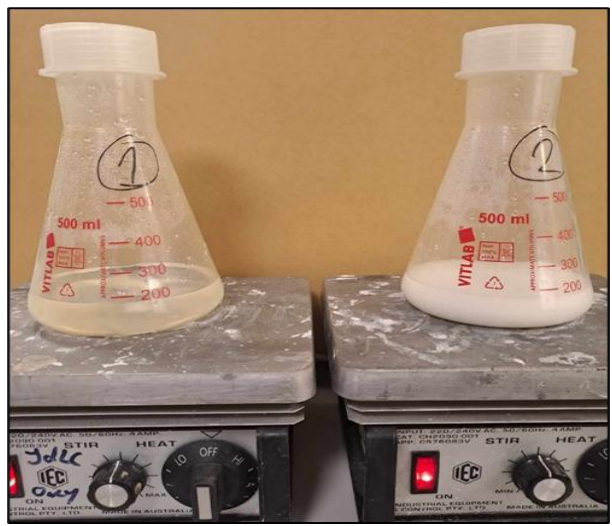

**Figure 8:** (a) An annotated ternary isotherm of the $H_2O$-$Na_2O$-$SiO_2$ system from Vail [28], highlighting composition regions referenced in the text. Green arrows indicate sequencing where deionised water, NaOH, and sodium silicate form a stable activator, whereas the red arrows show an alternative sequencing leading to precipitation. (b) Lab images from Skane et al. [16], illustrating the relatively stable (left) and unstable gel formed by the red-arrow-sequence method in the right flask.

## 3.4 Recommendations for Future Work

Further investigation using complementary analytical methods – such as Raman spectroscopy and Fourier Transform Infrared Spectroscopy (FTIR) with qNMR – is recommended to quantify both dissolved and solid silica phases, and to assess their respective impacts on geopolymer synthesis. This will improve understanding of total silica availability, guide more robust activator preparation protocols, and support quality control efforts in laboratory and industrial applications.

As a final recommendation, future work should experimentally compare activator solutions with identical molar compositions but different mixing histories e.g., one prepared and used within minutes, and another after a 24-hour "equilibration" period. Conceivably, the authors believe (and with preliminary support from literature [42]) that shorter preparation times which may result in enhanced reactivity and greater aluminosilicate dissolution from higher residual heat, potentially improving mechanical performance. Even if performance remains unchanged, these experiments may conceivably provide the same testing results as their controls which would allow for more practical and convenient usage of activator solutions with waiting times of 24 hours (in some literature cases) shortened to a period of minutes.

## 4 Conclusion

This study integrates experimentally validated silica speciation data, thermodynamic modelling, and solubility mapping of the $H_2O$-NaOH-$SiO_2$ system to improve the standardisation and optimisation of geopolymer activator solutions. The combined use of $^{29}Si$ NMR, solubility surface construction, and time–temperature modelling has yielded several key research findings:

1. **Minimised Preparation Times**: Results reinforce that activator solutions can be prepared and used within approximately 1–1.5 minutes, comprising ~1 minute for alkali thermodynamic stabilisation and seconds for silica speciation equilibration (the latter of which is supported from qNMR analyses in this paper). This significantly reduces the ≥ 24-hour preparation times commonly cited in the geopolymer literature, enabling faster and more efficient mixing procedures and process control.

2. **Quantification of Activator Thermochemical Stability Boundaries**: Activator solutions exhibit greater thermochemical stability and workability at elevated temperatures (below boiling point), while cooling increases the risk of irreversible precipitation. This transition into non-stable states can be quantified through the derived "time-instability window" for a given activator and visualised using three-dimensional solubility hypersurfaces. Minimising wasteful heat loss during mixing not only preserves solution stability but would also improve the dissolution of aluminosilicate precursors, enhancing overall reactivity and process effectiveness for eventual geopolymerisation.

3. **Optimisation of Activator Feedstock Sequencing**: Experimental results confirm that feedstock addition order affects stability. The optimal sequence is: (1) solvent water, (2) alkali hydroxide, and (3) soluble silicate. Altering this order increases the likelihood of entering unstable solubility regions and forming precipitates, even with identical feedstock compositions. Where conventional sequencing is not feasible (e.g., due to limited solvent capacity for hydroxide dissolution), this study provides tools to evaluate suitable alternative combinations and sequencing strategies.

Together, these findings challenge several common – but often undefined – practices in geopolymer literature and demonstrate that optimised activator solutions can be prepared rapidly, reproducibly, and with greater process control. Future work should explore the downstream effects of these optimised solutions on mechanical strength, durability and shrinkage with various aluminosilicate precursors and activator feedstocks. The mathematical model and stability mapping presented here provide a framework for developing standard operating procedures (SOPs) that support consistent, high-performance geopolymer synthesis across both laboratory and industrial applications.

# 5 Author Contributions

**Ramon Skane:** Conceptualisation, Methodology, Validation, Formal Analysis, Investigation, Project Administration, Software, Visualisation, Writing – Original Draft, Writing – Reviewing & Editing. **Franca Jones:** Validation, Resources, Writing – Review & Editing. **Arie van Riessen:** Funding Acquisition, Writing – Review & Editing. **Evan Jamieson:** Funding Acquisition, Writing – Review & Editing. **Xiao Sun:** Supervision,

Writing – Review & Editing. **William D. A. Rickard:** Validation, Supervision, Resources, Writing – Review & Editing.

## 6  Acknowledgements

The authors acknowledge the facilities at Curtin University, specifically the John de Laeter Centre for providing the laboratory for experimentation and storage of feedstocks, and to the staff at Coogee Chemicals Australia for assistance in procuring sodium silicate. Furthermore, the authors would like to extend an acknowledgement to Ching Yong Goh from the School of Molecular and Life Sciences at Curtin University for assistance in running and validating the NMR tests used in this study, and for the provision of calorimetry equipment, and finally to Liam Walsh for assistance in formulating some of the mathematics used for the experimentally validated mathematical model. This project has been undertaken as part of the Process Legacy subproject with financial support from the Future Battery Industries Cooperative Research Centre, which is established under the Australian Government's Cooperative Research Centres Program (Grant ID: 20180102).

## 7  Conflicts of Interest

The authors declare that they have no known competing financial interests or personal relationships that could have appeared to influence the work reported in this paper.

# 8 References

[1] J. Davidovits, "Geopolymers: Inorganic Polymeric New Materials," *Journal of Thermal Analysis,* vol. 37, pp. 429-441, 1991.

[2] L. Weng and Sagoe-Crentsil, "Dissolution processes, hydrolysis and condensation reactions during geopolymer synthesis: Part I - Low Si/Al Ratio Systems," *Journal of Material Sciences,* pp. 2997-3006, 2007.

[3] J. Aldred, J. Day and T. Glasby, "Geopolymer concrete—No longer labcrete," in *Proceedings of the 40th Conference on Our World in Concrete & Structures*, Singapore, 2015.

[4] T. Glasby, J. Day, R. Genrich and M. Kemp, "The Use of a Proprietary Geopolymer Concrete in Sewer Infrastructure," in *29th Biennial National Conference of the Concrete Institute of Australia*, Sydney, Australia, 2019.

[5] W. D. Rickard, R. Williams, J. Temuujin and A. Van Riessen, "Assessing the suitability of three Australian fly ashes as an aluminosilicate source for geopolymers in high temperature applications," *Materials Science and Engineering: A,* vol. 528, no. 9, pp. 3390-3397, 2011.

[6] B. C. McLellan, R. P. Williams, J. Lay, A. van Riessen and G. D. Corder, "Costs and carbon emissions for geopolymer pastes in comparison to ordinary portland cement," *Journal of Cleaner Production,* pp. 1080-1090, 2011.

[7] E. Jamieson, B. McLellan, A. van Riessen and H. Nikraz, "Comparison of embodied energies of Ordinary Portland Cement with Bayer-derived geopolymer products," *Journal of Cleaner Production,* vol. 99, pp. 112-118, 2015.

[8] W. M. Kriven, C. Leonelli, J. L. Provis, A. R. Boccaccini, C. Attwell, V. S. Ducman, C. Ferone, S. Rossignol, T. Luukkonen, J. S. J. van Deventer, J. V. Emiliano and J. E. Lombardi, "Why geopolymers and alkali-activated materials are key components of a sustainable world: A perspective contribution," *Journal of the Americal Ceramic Society,* vol. 107, no. 8, pp. 5159-5177, 2024.

[9] T. How Tan, K. H. Mo, T.-C. Ling and S. Hin Lai, "Current development of geopolymer as alternative adsorbent for heavy metal removal," *Environmental Technology & Innovation,* vol. 18, p. 100684, 2020.

[10] A. van Riessen, E. Jamieson, C. S. Kealley, R. D. Hart and R. P. Williams, "Bayer-geopolymers: An exploration of synergy between the alumina and geopolymer industries," *Cement and Concrete Composites,* vol. 41, pp. 29-33, 2013.

[11] Beyond Zero Emissions, "Rethinking Cement," Beyond Zero Emissions Inc., Melbourne, 2017.

[12] O. F. Arbeláez-Pérez, V. Senior-Arrieta, J. Hernán Gómez Ospina, S. Herrera Herrera, C. Ferney Rodríguez Rojas and A. María Santis Navarro, "Carbon dioxide emissions from traditional and modified concrete. A review," *Environmental Development,* vol. 52, p. 101036, 2024.

[13] J. Matsimbe, M. Dinka, D. Olukanni and I. Musonda, "A Bibliometric Analysis of Research Trends in Geopolymer," *Materials,* vol. 15, pp. 1-20, 2022.

[14] G. D. Poggetto, C. Leonelli and A. Spinella, "Influence of anionic silica forms in clear sodium silicate precursors on metakaolin geopolymerisation via 29Si and 27Al MAS-NMR and microstructural studies," *J Mater Sci,* vol. 59, pp. 16963-16980, 2024.

[15] J. Matsimbe, M. Dinka, D. Olukanni and I. Musonda, "Geopolymer: A Systematic Review of Methodologies," *Materials,* vol. 15, no. 19, p. 6852, 2022.

[16] R. Skane, P. A. Schneider, F. Jones, A. van Riessen, E. Jamieson, X. Sun and W. D. A. Rickard, "Predicting the stability of geopolymer activator solutions for optimised synthesis through thermodynamic modelling," *Chemical Engineering Journal,* vol. 515, p. 163543, 2025.

[17] D. Dimas, I. Giannopoulou and Panias, "Polymerization in sodium silicate solutions: a fundamental process in geopolymerization technology," *Jounral of Material Sciences,* pp. 3719-3730, 2009.

[18] J. L. Provis, "Activating solution chemistry for geopolymers," in *Geopolymers: Structures, Processing, Properties and Industrial Applications*, J. L. Provis and J. S. J. van Deventer, Eds., Woodhead Publishing Series in Civil and Structural Engineering, 2009, pp. 50-71.

[19] I. Halasz, M. Agarwal, L. Runbo and N. Miller, "What can vibrational spectroscopy tell about the structure of dissolved sodium silicates?," *Microporous and Mesoporous Materials,* vol. 135, pp. 74-81, 2010.

[20] M. Matinfar and J. A. Nychka, "A review of sodium silicate solutions: Structure, gelation, and syneresis," *Advances in Colloid and Interface Science,* vol. 322, p. 103036, 2023.

[21] J. Provis, P. Duxson, G. C. Lukey, F. Separovic, W. M. Kriven and J. S. J. van Deventer, "Modeling Speciation in Highly Concentrated Alkaline Silicate Solutions," *Industrial & Engineering Chemistry Research,* vol. 44, no. 23, pp. 8899-8908, 2005.

[22] T. W. Swaddle, J. Salerno and P. A. Tregolan, "Aqueous aluminates, silicates, and aluminosilicates," *Chemical Society Reviews,* vol. 23, no. 5, pp. 319-325, 1994.

[23] G. Engelhardt, D. Zeigan, H. Jancke, D. Hoebbel and W. Wieker, "29Si NMR Spectroscopy of Silicate Solutions. II. On the Dependence of Structure of Silicate Anions in Water Solutions from the Na:Si Ratio," *Z. Anorg. Allg. Chem,* vol. 17, p. 418, 1975.

[24] L. Vidal, E. Joussein, M. Colas, J. Cornette, J. Sanz, I. Sobrados, J.-L. Gelet, J. Absi and S. Rossignol, "Controlling the reactivity of silicate solutions: A FTIR, Raman and NMR Study," *Colloids and Surfaces A: Physicochemical and Engineering Aspects,* vol. 503, pp. 101-109, 2016.

[25] X. Yang, W. Zhu and Q. Yang, "The Viscosity Properties of Sodium Silicate Solutions," *J Solution Chem,* vol. 37, pp. 73-83, 2008.

[26] H. Jansson, D. Bernin and K. Ramser, "Silicate species of water glass and insights for alkali-activated green cement," *AIP Advances,* vol. 5, no. 6, p. 067167, 2015.

[27] M. Zhao, G. Liu, C. Zhang, W. Guo and Q. Luo, "State-of-the-Art of Colloidal Silica-Based Soil Liquefaction Mitigation: An Emerging Technique for Ground Improvement," *Appl. Sci.,* vol. 10, no. 1, p. 15, 2020.

[28] J. G. Vail, "Soluble Silicates: Their Properties and Uses," *American Chemical Society. Monograph Series,* no. 116, 1952.

[29] P. W. Brown, "The System Na2O-CaO-SiO2-H2O," *J. Am. Ceram. Soc,* vol. 73, no. 11, pp. 3457-61, 1990.

[30] J. D. Rimstidt, "Quartz solubility at low temperatures," *Geochimica et Cosmochimica Acta,* vol. 61, pp. 2553-2558, 1997.

[31] F-Chart Software, "Engineering Equation Solver," F-Chart Software, 2025. [Online]. Available: https://fchartsoftware.com/ees/.

[32] R. K. Harris, E. K. F. Bahlmann, K. Metcalfe and E. G. Smith, "Quantitative silicon-29 NMR investigations of highly concentrated high-ratio sodium silicate solutions," *Magnetic Resonance in Chemistry,* vol. 31, pp. 743-747, 1993.

[33] S. D. Kinrade and T. W. Swaddle, "Silicon-29 NMR studies of aqueous silicate solutions. 1. Chemical shifts and equilibria," *Inorganic Chemistry,* vol. 27, no. 23, pp. 4081-4332, 1988.

[34] P. Sipos, A. Stanley, S. Bevis, G. Hefter and P. M. May, "Viscosities and Densities of Concentrated Aqueous NaOH/NaAl(OH)4 Mixtures at 25 °C," *J. Chem. Eng,* vol. 46, pp. 657-661, 2001.

[35] A. Lach, L. André, A. Lassin, M. Azaroual, J.-P. Serin and P. Cézac, "A New Pitzer Parameterization for the Binary NaOH–H2O and Ternary NaOH–NaCl–H2O and NaOH–LiOH–H2O Systems up to NaOH Solid Salt Saturation, from 273.15 to 523.15 K and at Saturated Vapor Pressure," *Journal of Solution Chemistry,* vol. 44, pp. 1424-1451, 2015.

[36] W. M. Haynes, D. R. Lide and T. J. Bruno, Eds., "Physical Constants of Inorganic Compounds," in *CRC Handbook of Chemistry and Physics*, 97th ed., Boca Raton, Florida: CRC Press, 2017, p. 4.86.

[37] P. Duxson, J. L. Provis, G. C. Lukey, S. W. Mallicoat, W. M. Kriven and J. S. van Deventer, "Understanding the relationship between geopolymer composition, microstructure and mechanical properties," *Colloids and Surfaces A: Physicochemical and Engineering Aspects,* vol. 269, no. 1-3, pp. 47-58, 2005.

[38] E. K. F. Bahlmann, R. K. Harris, K. Metcalfe, J. W. Rockliffe and E. G. Smith, "Silicon-29 NMR self-diffusion and chemical-exchange studies of concentrated sodium silicate solutions," *J. Chem. Soc., Faraday Trans.,* vol. 93, no. 1, pp. 93-98, 1997.

[39] J. Provis and J. S. J. van Deventer, "Geopolymerisation kinetics. 2. Reaction kinetic modelling," *Chemical Engineering Science,* vol. 62, pp. 2318-2329, 2007.

[40] G. L. Hovis, M. J. Toplis and P. Richet, "Thermodynamic mixing properties of sodium silicate liquids and implications for liquid–liquid immiscibility," *Chemical Geology,* vol. 213, no. 1-3, pp. 173-186, 2004.

[41] J. Nordström, E. Nilsson, P. Jarvol, M. Nayeri, A. Palmqvist, J. Bergenholtz and A. Matic, "Concentration- and pH-dependence of highly alkaline sodium silicate solutions," *Journal of Colloid and Interface Science,* vol. 356, no. 1, pp. 37-45, 2011.

[42] C. Kuenzel and N. Ranjbar, "Dissolution mechanism of fly ash to quantify the reactive aluminosilicates in geopolymerisation," *Resources, Conservation & Recycling,* 2019.
## 9 Supplementary Information & Appendices

### 9.1 Appendix A: List of Variables and Notation

Refer below to the nomenclature used throughout this document and from Skane et al [16] where the following subscripts are used but omitted from the list below for conciseness:

- *i* and *f* are used with some symbols to indicate the initial and final states of those variables (e.g. $C_{NaOH,i}$ and $C_{NaOH,f}$).
- *sys* and *j* are used to characterise the full system and its sub-components respectively.
- *x* to denote different feedstocks (i.e. NaOH pellets, solvent water or sodium silicate).
- *MS* and *US*, which respectively represent metastable and unstable stability states, are used as subscripts with some symbols to indicate their stability state.
- Values are added for constant variables used in the model (i.e. initial conditions).

| Symbol | Parameter Definition | [Units] |
|---|---|---|
| $a_k$ | Polynomial coefficient of number *k* | *(varies)* |
| $C_{NaOH}$ | Concentration of NaOH | *mol/L* |
| $C_{SiO2}$ | Concentration of SiO$_2$ | *mol/L* |
| $m_x$ | Mass of feedstock x | *g* |
| $n_d$ | Polynomial degree of number *k* | *(varies)* |
| $Q^n$ | Silica species with *n* bridging oxygen atoms | *[wt.%]* |
| $Q^{nc}$ | Cyclic geometry silica species with *n* bridging oxygen atoms | *[wt.%]* |
| $Q^{nsq}$ | Square geometry silica species with *n* bridging oxygen atoms | *[wt.%]* |
| $Q_T^n$ | Silica species with *n* bridging oxygen atoms as a multivariable function of temperature | *[wt.%]* |
| $Q_{Tot}^n$ | Summated silica species with *n* bridging oxygen atoms of all geometries | *[wt.%]* |
| $T_{Amb}$ | Ambient Temperature | °C |
| $T_{\lambda MS}$ | Activator solution metastable temperature | °C |
| $T_{Ref}$ | Reference Temperature | °C |
| $T_{Soln}$ | Temperature of activator solution at time *t* | °C |
| $T_{Stable}$ | Initial stability temperature of activator solution after all feedstocks added | °C |
| $t_{stable}$ | Stability Time of Activator solution after all feedstocks added | *s* |
| $T_{\lambda US}$ | Activator solution unstable temperature | °C |
| $z_i$ | Charge number *i* for species Q$^n$ | - |
| $\lambda_{MS}$ | Metastable stability vector | - |
| $\lambda_{US}$ | Unstable stability vector | - |
| $\omega(t)_{SS}$ | Context piecewise function for sodium silicate feedstock addition | - |

## 9.2 Appendix B: Extended Modelling Equations

Conventional relationships such as density equations (mass/volume), weight and molar fractions (etc.) were used throughout the model. From the general Equation X in section 2.2, the functions fitted for modelling equations are observable in the tables and equations in the subsections below.

### 9.2.1 Solubility Modelling Equations

Table B.1 Parameter for solubility model equations from this work and other literature sources.

| Equation ID | Variable | | Coefficient (*a*) to Degree *n* | | | | | | |
|---|---|---|---|---|---|---|---|---|---|
| | Dependent | Independent | *0* | *1* | *2* | *3* | *4* | *5* | *6* |
| B.1 | $C_{MS}$ | $C_{SiO2}$ | 8.98E+00 | -1.22E+01 | 8.57E+00 | -2.68E+00 | 4.30E-01 | -3.00E-02 | 1.00E-03 |
| B.2 | $C_{US} = C_{NaOH}$ | $T_{Soln}$ | 1.10E+01 | 9.28E-01 | -3.57E-02 | 9.63E-04 | -1.01E-05 | 4.53E-08 | -7.38E-11 |
| B.3 | $\partial C_{US}/\partial T$ | $T_{Soln}$ | 9.28E-01 | -7.13E-02 | 2.88E-03 | -4.00E-05 | 2.26E-07 | -4.43E-10 | - |
| B.4 | $C_{MS} = C_{SiO2}$ | $T_{Soln}$ | 9.10E-05 | 1.72E-06 | 5.42E-08 | 2.07E-10 | - | - | - |
| B.5 | $\partial C_{MS}/\partial T$ | $T_{Soln}$ | 1.72E-06 | 1.08E-07 | 6.20E-10 | - | - | - | - |

Table B.2 Equation B domains, units and references.

| Equation ID | Domain | | | Reference |
|---|---|---|---|---|
| | Min | Max | Units | |
| B.1 | 0 | 8 | Mol/L | |
| B.2 | 15 | 95 | °C | [35] |
| B.3 | 15 | 95 | °C | |
| B.4 | 0 | 200 | °C | [30] |
| B.5 | 0 | 200 | °C | |

$$\frac{\partial C_{US}}{\partial C_{SiO_2}}(C_{SiO2}) = \begin{cases} -4.92, & C_{SiO_2} \leq 3.47 \\ 1.95, & C_{SiO_2} > 3.47 \end{cases} \quad (B.1)$$

### 9.2.2 Chemical Speciation Modelling Equations

Table C.1. Parameter for silicate speciation polynomial fitted model equations as a function of {Molar SiO$_2$/Na$_2$O ∈ [0.25, 4]} from combined literature sources.

| Equation ID | Dependent Variable | Coefficient (*a*) to Degree *n* | | | | | | | Reference |
|---|---|---|---|---|---|---|---|---|---|
| | | *0* | *1* | *2* | *3* | *4* | *5* | *6* | |
| C.1 | $Q^0$ | 4.60E-01 | 1.27E+00 | -1.92E+00 | 1.49E-01 | 5.51E-01 | -1.65E-01 | - | |
| C.2 | $Q^1$ | -6.11E-01 | 7.86E+00 | -2.27E+01 | 3.02E+01 | -2.03E+01 | 6.66E+00 | -8.47E-01 | |
| C.3 | $Q^2$ | 6.65E-01 | -5.36E+00 | 1.48E+01 | -1.89E+01 | 1.31E+01 | -4.69E+00 | 6.70E-01 | [21] |
| C.4 | $Q^3$ | 3.85E-01 | -2.89E+00 | 7.05E+00 | -7.33E+00 | 3.49E+00 | -6.07E-01 | - | |
| C.5 | $Q^4$ | N/A | N/A | N/A | N/A | N/A | N/A | N/A | |
| C.6 | $Q^0$ | N/A | N/A | N/A | N/A | N/A | N/A | N/A | |
| C.7 | $Q^1$ | 1.44E+00 | -1.37E+00 | 4.96E-01 | -7.47E-02 | 3.59E-03 | - | - | |
| C.8 | $Q^2$ | -8.98E-01 | 2.68E+00 | -1.67E+00 | 4.10E-01 | -3.57E-02 | - | - | [32] |
| C.9 | $Q^3$ | -1.52E+00 | 1.81E+00 | -5.39E-01 | 5.38E-02 | - | - | - | |
| C.10 | $Q^4$ | 3.74E+00 | -5.61E+00 | 3.01E+00 | -6.83E-01 | 5.66E-02 | - | - | |

Table C.2. Parameter for silicate speciation polynomial fitted model equations as a function of {T$_{Soln}$ ∈ [10, 77.5] °C} adapted from one-sided backwards finite difference method of results from Kinrade and Swaddle [].

| Equation ID | Coefficient (*a*) to Degree *n* |
|---|---|

| | Dependent Variable | 0 | 1 | 2 | 3 |
|---|---|---|---|---|---|
| C.11 | $\partial Q^0/\partial T$ | -2.90E+00 | 1.33E-01 | -7.45E-04 | 4.14E-06 |
| C.12 | $\partial Q^1/\partial T$ | -4.81E+00 | 2.24E-01 | -1.43E-03 | 6.56E-06 |
| C.13 | $\partial Q^2/\partial T$ | -8.51E+00 | 4.69E-01 | -5.45E-03 | 2.11E-05 |
| C.14 | $\partial Q^3/\partial T$ | 1.62E+01 | -8.26E-01 | 7.63E-03 | -3.18E-05 |
| C.15 | $\partial Q^4/\partial T$ | N/A | N/A | N/A | N/A |

## 9.3  Appendix C: Combined Speciation Model Reference

The following models from literature were adapted into Figure 4 within the main text to predict speciation changes as a function of both the activator solution's molar $SiO_2/Na_2O$ and temperature.

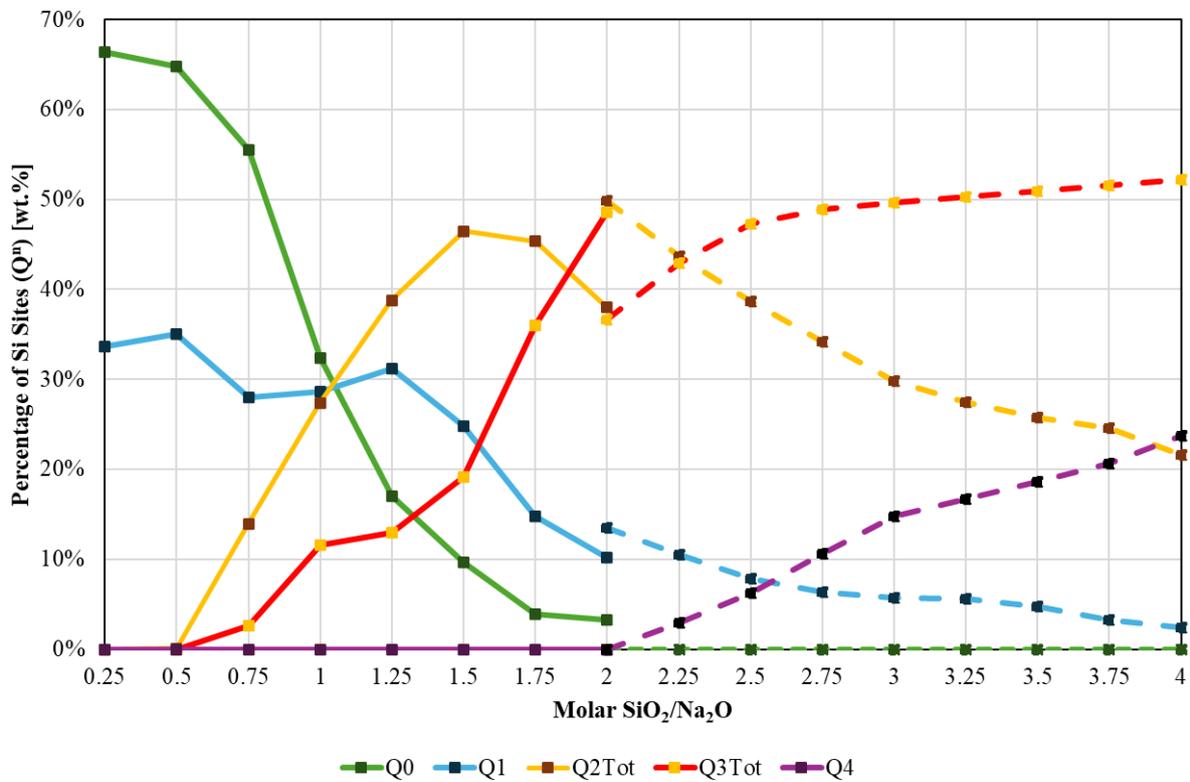

**Figure C.1**: Silicate Speciation in activator solutions as a function of its molar $SiO_2/Na_2O$ obtained by $^{29}Si$ NMR spectroscopy with solid and dashed lines adapted from the works of Provis et al [21] and Harris et al [32] respectively.

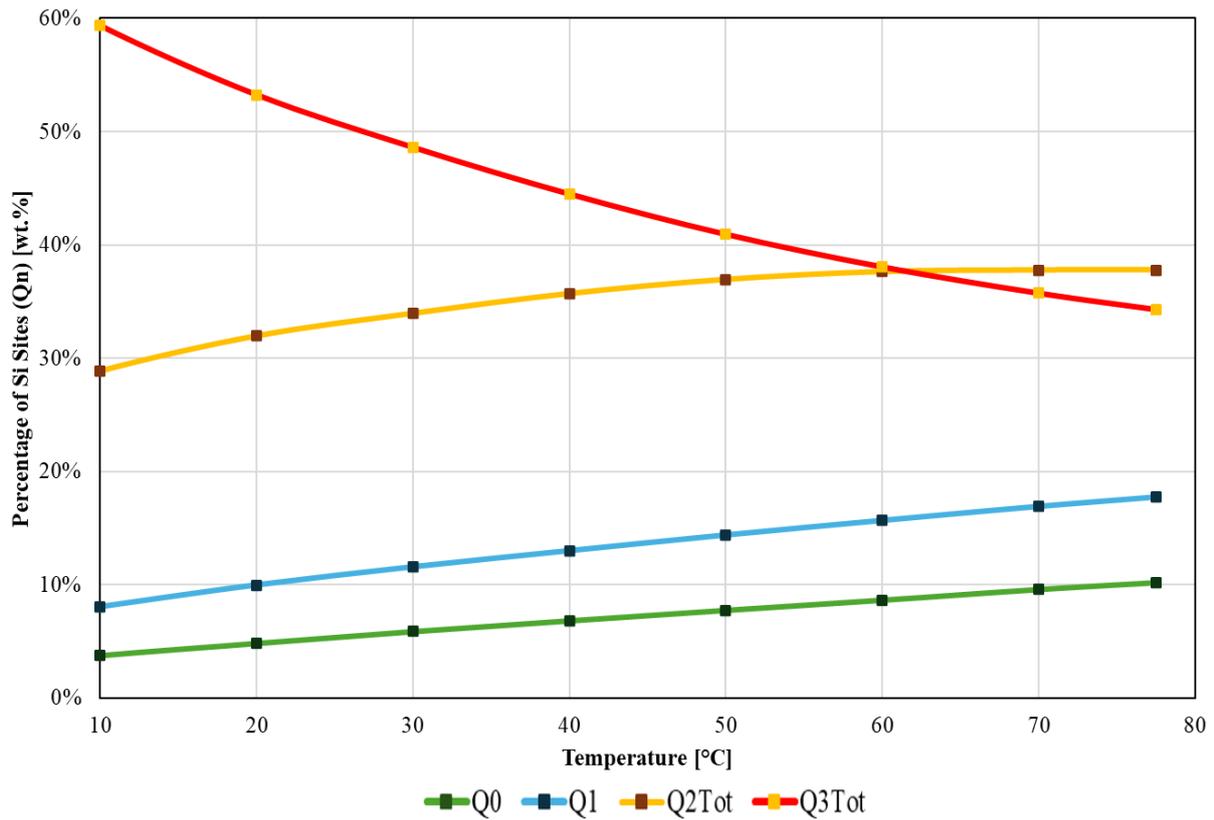

**Figure C.2**: Silicate Speciation in activator solutions as a function of solution temperature obtained by $^{29}$Si NMR spectroscopy adapted from the works of Kinrade and Swaddle [33].

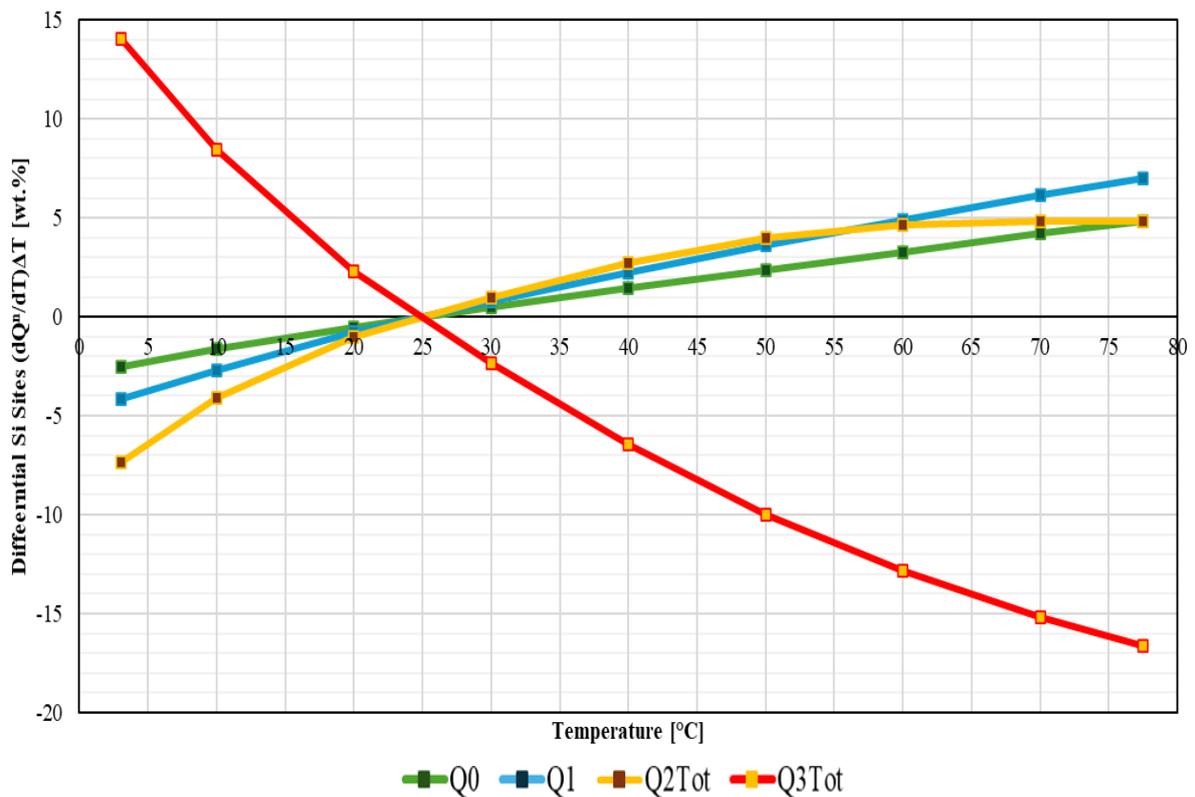

**Figure C.3**: Silicate Speciation as a function of temperature with respect to a reference temperature of 25 C as adapted from Figure C.2.

## 9.4 Appendix D: Metastable Solubility Solution Space

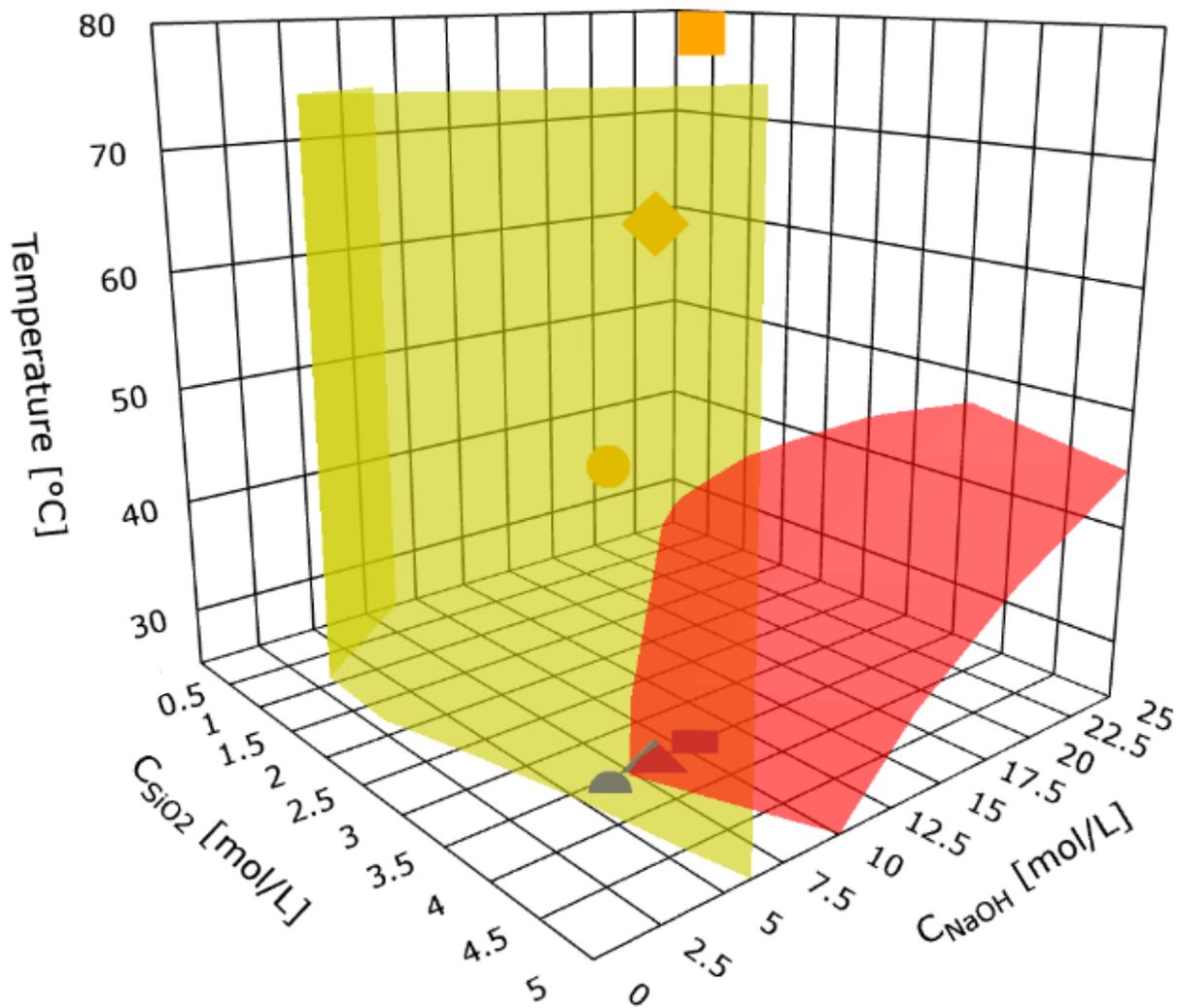

**Figure D.1**: Activator solution stability solution space conveying metastable ("MS", yellow) and unstable ("US", red) hypersurfaces. Centred dot points representing the A, C and E experimental solutions plotted as squares, diamonds and circles where the red and blue coloured icons represent the temperatures at sodium silicate stabilisation and 25 °C respectively.